\documentclass[fleqn,usenatbib]{mnras}

\DeclareRobustCommand{\VAN}[3]{#2}
\let\VANthebibliography\thebibliography
\def\thebibliography{\DeclareRobustCommand{\VAN}[3]{##3}\VANthebibliography}

\usepackage{graphicx}	
\usepackage{amsmath}	
\usepackage{amssymb}	
\usepackage{threeparttable}
\usepackage[T1]{fontenc}
\usepackage{newtxtext,newtxmath}

\usepackage{booktabs,caption}

\title{Capture Rate of Weakly Interacting Massive Particles (WIMPs) In Binary Star Systems}

\author[Ebrahim Hassani]{
Ebrahim Hassani,$^{1}$\thanks{E-mail: ebrahim.hassani@birjand.ac.ir}
Hossein Ebadi,$^{2}$
Reza Pazhouhesh$^{1}$
and Mohammad Hosseinirad$^{3}$
\\
$^{1}$  Department of Physics, Faculty of Sciences, University of Birjand, Birjand, Iran \\
$^{2}$  Department of Theoretical Physics and Astrophysics, Physics Faculty, University of Tabriz, PO Box 51664, Tabriz, Iran \\
$^{3}$  School of Astronomy, Institute for Research in Fundamental, Sciences (IPM), Tehran, Iran \\
}

\date{Accepted XXX. Received YYY; in original form ZZZ}

\pubyear{2020}

\begin{document}
\label{firstpage}
\pagerange{\pageref{firstpage}--\pageref{lastpage}}
\maketitle

\begin{abstract}
The distribution of dark matter (DM) inside galaxies is not uniform. Near the central regions, its density is the highest. Then, it is logical to suppose that, inside galaxies, DM affects the physics of stars in central regions more than outer regions. Besides, current stellar evolutionary models did not consider DM effects in their assumptions. To consider DM effects, at first one must estimate how much DM a star contains. The capture rate (CR) of DM particles by individual stars was investigated already in the literature. In this work, we discuss how CR can be affected when stars are members of binary star systems (BSS) (instead of studying them individually). When a star is a member of a BSS, its speed changes periodically due to the elliptical motion around its companion star. In this work, we investigated CR by BSSs in different BSS configurations. In the end, we discussed observational signatures that can be attributed to the DM effects in BSSs.
\end{abstract}

\begin{keywords}
Dark matter, Binary stars, Milky way galaxy
\end{keywords}

\section{Introduction} \label{Introduction}
According to the standard model of cosmology ($ \Lambda CDM $ model), about 25 percent of the matter in the universe is in the form of dark matter (DM) \citep{cosmology_weinberg} . Besides, many other observational evidences support the existence of DM in large and small scale structures (e.g. rotation curves of galaxies \citep{Sofue2001} , simulations of galaxies \citep{KUHLEN201250}, and simulation of the universe \citep{Pillepich2018}) . Rotation curves of galaxies show that DM distributed non-uniformly \citep{1996ApJ...462..563N} inside galaxies. Then, we can say, stars evolve inside galaxies while they are immersed in the DM. Therefore, DM must affect the evolutionary course of stars inside galaxies \citep{2008PhRvD..77d7301F,2008dmap.conf..387S,2009MNRAS.394...82S} .
Signs of DM effects on stars were investigated before this study in the literature. For example: 
\begin{itemize}
\item For the first time Steigman used DM supposition on the sun to solve the discrepancy between the observed and calculated solar neutrino fluxes \citep{1978AJ.....83.1050S} . Since then, many studies had conducted to solve the solar neutrino problem using the supposition that DM particles annihilate inside the sun.

\item Simulation of dwarf galaxies, with the same mass, shows that the halo of DM around evolved dwarf galaxies can be heated-up by star formation process inside galaxies and then push the DM around \citep{Read_2018fxs} . The more evolved the dwarf galaxy is then, the more DM halo heated-up by stars.

\item Stars that evolve near the Galactic massive black hole show signs of young and old stars simultaneously, which is known as the paradox of youth problem. Supposing that DM particles annihilate inside stars then it is possible to solve this problem \citep{Hassani_2020zvz} .

\item In addition to the normal stars, the effects of DM on compact stars (white dwarfs and neutron stars) were also investigated in the literature. For instance, the annihilation of DM particles inside compact stars can flatter out their temperature or it is possible to constrain DM properties using compact stars \citep{2008PhRvD..77b3006K,2010PhRvD..82f3531K, Bertone_2008, doi:10.1142/S0218271819500020, REZAEI20181, Rezaei_2017}.
\end{itemize}

According to the definition, capture rate (CR) of DM particles by a round massive body (like Earth, Sun, neutron stars, etc) is the number of DM particles that are gravitationally bound to that body by passing the time \citep{Gould_1987ir} . For the first time, Press and Spergel calculated the CR of weakly interacting massive particles (WIMP) by the sun \citep{1985ApJ...296..679P} . Then, Gould generalized the CR relation for other round objects (like planets and stars) \citep{Gould_1987ir}. Since then, many other studies used Gould relation to calculate CR by round massive bodies \citep{Bell:2020jou, 2020arXiv200405312L, 2019JCAP...06..054B, 2018JCAP...09..018B, 2019JCAP...12..043N, 2018JHEP...04..074C, 2017JCAP...01..059C, 2016PhRvD..93a5014F, 2014PhRvD..89b5003L, 2020PhLB..80435403G, 2019JCAP...08..018D, 2018PhRvD..98k5027G}. In this study, we used Gould relation to calculate CR by stars (see section \ref{DM_CR_by_stars} for more details).

Accumulation of DM particles inside massive bodies (weather they annihilate or they do not) can alter the structure and evolutionary course of stars \citep{2008dmap.conf..387S,2009MNRAS.394...82S,2008PhRvD..77d7301F}. Therefore, they can be responsible for some observational phenomenon like gamma ray emission \citep{2017PhRvD..95e5031B} and neutrino emission from stars \citep{2015APh....62...12A}. This effect is boosted for stars that are located in high DM density environments, like near the Galactic massive black hole .Then, it is important to estimate the exact value of CR by massive bodies as much as possible. CR for different kind of round massive bodies like the Moon \citep{2020PhLB..80435403G, 2020PhRvD.102b3024C} , planets (like Earth and exoplanets) \citep{2019JHEP...05..039T, 2017JCAP...01..059C, ADLER2009203, Iorio:2010gh}, the Sun \citep{2019JCAP...12..043N, Widmark:2017bzr, ISI:000413582300008}, other stars \citep{Lopes2011, 2008dmap.conf..387S, 2019JCAP...12..051I, 2017PhRvD..96f3002B, 2008PhRvD..78l3510T, 2009MNRAS.394...82S,2009ApJ...705..135C}, compact stars \citep{Bertone_2008, 2010PhRvD..81j3531H, 2011PhRvD..83h3512K,  2010PhRvD..82f3531K} are estimated in the literature.

The effects of DM on compact binary systems were investigated in the literature too
\citep{Hassani2020f, 2015PhRvD..92l3530P, PhysRevD.96.063001, GOMEZ2019100343, CAPUTO20181}. But, to the best of our knowledge, there is no similar topic for normal 
(non-compact) binary systems. So, in this study, we estimated the CR by binary star systems (BSSs) and then discussed the effects of binary parameters on CR. Section \ref{theories_and_models} is devoted to the theories and models that are used in this work. The formulas that are used in this study, is derived in this section. In Section \ref{binary_system_parameters}, the effects of BSS parameters 
on CR were investigated. Finally, Section \ref{conclusion_discussion} is devoted to conclusions and discussions. Possible observational signs of DM effects in BSSs are discussed in this section too.

\section{Theories and models} \label{theories_and_models}

\subsection{Capture rate by stars} \label{DM_CR_by_stars} 
We used Gould relations to calculate CR by stars \citep{Gould_1987ir}. Total CR by different elements inside stars can be calculated using Gould relation:

\begin{equation} \label{total_cap_rate}
C_{\chi}(t) = \sum_{i}\int_{0}^{R_{\ast}} 4 \pi r^{2}\int_{0}^{\infty} \frac{f_{v_{\ast}}(u)}{u}\omega \Omega_{v,i}^{-} (\omega)du dr .
\end{equation}

It is worth mentioning that recently a more generalized form of CR was developed in the paper \citep{Dasgupta2020}. Authors considered arbitrary mass mediators in the CR relation. But, for the purposes of this paper, there is not a significant difference between the Gould relation and the relation developed in that paper \citep{Dasgupta2020}.
In sections \ref{S_CR_by_hydrogen} and \ref{Sec_CR_by_hevier} we calculated CR relation for hydrogen and heavier elements separately. In equation \ref{total_cap_rate} , $\Omega_{v,i}^{-}$ is the rate at which a WIMP with velocity $ \omega $ scatters to a velocity less than $v$ (escape velocity from the surface of the star) and then gravitationally bounds. For hydrogen atoms $\Omega_{v,i}^{-}$ is:

\begin{equation} \label{omega_hydrogen}
\Omega_{v,H}^{-} (\omega) = \frac{\sigma_{\chi,H}n_{H}(r)}{\omega}(v_{e}^{2}-\frac{\mu_{-,H}^{2}}{\mu_{H}}u^{2}) \theta (v_{e}^{2}-\frac{\mu_{-,H}^{2}}{\mu_{H}}u^{2})
\end{equation}

where $\theta$ is the step function. For heavier elements $ \Omega_{v,i}^{-} (\omega) $ is :

\begin{multline} \label{omega_hevier}
\Omega_{v,i}^{-} (\omega) = \frac{\sigma_{\chi ,i} n_{i}(r)}{\omega} \frac{2 E_{0}}{m_{\chi}} \frac{\mu^{2}_{+,i}}{\mu_{i}} \times \\ 
\left \{ exp(-\frac{m_{\chi } u^{2}}{2E_{0}})-exp(-\frac{m_{\chi} u^{2}}{2E_{0}}\frac{\mu_{i}}{\mu^{2}_{+,i}}) exp(-\frac{m_{\chi}v_{e}^{2}}{2E_{0}}\frac{\mu_{i}}{\mu^{2}_{-,i}}(1-\frac{\mu_{i}}{\mu^{2}_{+,i}})) \right \}
\end{multline}

in which $E_{0}$ is the characteristic coherence energy and can be calculated using (see reference \citep{Gould_1987ir} for more details):

\begin{equation} \label{coherence_energy}
E_{0} = \frac{3 \hbar^{2}}{2 m_{n,i}(0.91 m_{n,i}^{1/3}+0.3)^{2}}
\end{equation}

In equation \ref{total_cap_rate} we have:

\begin{equation}
\mu_{\mp,i} \equiv \frac{\mu_{i}\mp1}{2}
\end{equation}
and
\begin{equation}
\mu_{i} \equiv \frac{m_{\chi}}{m_{n,i}}
\end{equation}

$f_{v_{\ast}}(u)$ is the velocity distribution function of DM particles at the location of the star. $f_{v_{\ast}}(u)$ usually considered to be a Maxwell-Boltzmanian distribution \citep{2009MNRAS.394...82S} with a dispersion velocity $ \overline{v}_{\chi} $ :

\begin{equation} \label{velocity_distribution}
f_{v,\ast}(u) = f_{0}(u) exp (-\frac{3v^{2}_{\ast}}{2\overline{v}^{2}_{\chi }})\frac{sinh(3uv_{\ast }/\overline{v}^{2}_{\chi})}{3uv_{\ast}/\overline{v}^{2}_{\chi}}
\end{equation}

in which $ f_{0}(u) $ is the velocity dispersion of the DM particles in the halo and is:

\begin{equation} \label{velocity_distru_halo}
f_{0}(u) = \frac{\rho_{\chi}}{m_{\chi}}\frac{4}{\sqrt{\pi}}(\frac{3}{2})^{3/2}\frac{u^{2}}{\overline{v}^{3}_{\chi}} exp (-\frac{3u^{2}}{2\overline{v}^{2}_{\chi}})
\end{equation}

 in which $\rho_{\chi}$ is the DM density around BSS and $ \sigma_{\chi,i} $ is the scattering cross section from an element $i$. For hydrogen atoms , $ \sigma_{\chi,i} $  is:

\begin{equation} \label{scattering_hydrogen}
\sigma_{\chi,H} = \sigma_{\chi,SI} + \sigma_{\chi,SD}
\end{equation}

and for elements heavier than hydrogen it is :

\begin{equation} \label{scattering_cross_section}
\sigma_{\chi,i} = \sigma_{\chi,SI} A_{i}^{2} (\frac{m_{\chi}m_{n,i}}{m_{\chi}+m_{n,i}})^{2}(\frac{m_{\chi}+m_{p}}{m_{\chi}m_{p}})^{2}
\end{equation}

In above equations
$\sigma_{\chi,SI}$ is the spin-independent DM-nucleon scattering cross section,
$\sigma_{\chi,SD}$ is the spin-dependent DM-nucleon scattering cross section,
$m_{\chi}$ is the mass of the DM particles (WIMPs, in the case of this study), 
$m_{n,i}$ is the nuclear mass of the element $i$,
$A_{i}$ is the atomic number of the element $i$,
$n_{i}(r)$ is the number density of the element $i$ at a radius $r$ from the center of the star,
and $R_{\ast}$ is the radius of the star.

In the coming two sections, we will calculate CR relation for hydrogen and heavier elements separately.
\subsubsection{Capture rate by hydrogen atoms} \label{S_CR_by_hydrogen}
After putting equations \ref{omega_hydrogen} , \ref{velocity_distribution} , \ref{velocity_distru_halo} and \ref{scattering_hydrogen} into equation \ref{total_cap_rate} and then some arrangements, we obtain the CR relation for hydrogen atoms:

\begin{multline} \label{CR_by_hydrogen}
C_{\chi ,H} = \left [ 4\sqrt{6\pi } \frac{\rho_{\chi}}{m_{\chi}} \frac{1}{\overline{v}_{\chi }v_{\ast}} exp(-\frac{3v^{2}_{\ast}}{2\overline{v}^{2}_{\chi}}) \right ] \times \\
\left [ \sigma_{\chi,SI} + \sigma_{\chi,SD} \right ] 
\left [ \int_{0}^{R_{\ast}} n_{H}(r) r^{2} dr \right ] \times \\
\left [ \int_{0}^{\infty } exp(-\frac{3u^{2}}{2\overline{v}^{2}_{\chi}}) sinh(\frac{3uv_{\ast}}{\overline{v}^{2}_{\chi}}) (v_{e}^{2}-\frac{\mu_{-,H}^{2}}{\mu_{H}}u^{2}) \theta (v_{e}^{2}-\frac{\mu_{-,H}^{2}}{\mu_{H}}u^{2}) du \right ]
\end{multline}

\subsubsection{Capture rate by heavier elements} \label{Sec_CR_by_hevier}
After putting equations \ref{omega_hevier} , \ref{velocity_distribution} , \ref{velocity_distru_halo} and \ref{scattering_cross_section} into equation \ref{total_cap_rate} and then some arrangements, we obtain the CR relation for heavier elements:

\begin{multline} \label{CR_by_hevier}
C_{\chi ,i} = \left [ 8\sqrt{6\pi } \frac{\rho_{\chi}}{m_{\chi}^{2}} \frac{E_{0}}{\overline{v}_{\chi }v_{\ast}} \frac{\mu^{2}_{+,i}}{\mu_{i}} exp(-\frac{3v^{2}_{\ast}}{2\overline{v}^{2}_{\chi}}) \right ] \times \\
\left [ \sigma_{\chi,SI} A_{i}^{2} (\frac{m_{\chi}m_{n,i}}{m_{\chi}+m_{n,i}})^{2}(\frac{m_{\chi}+m_{p}}{m_{\chi}m_{p}})^{2} \right ]
\left [ \int_{0}^{R_{\ast}} n_{H}(r) r^{2} dr \right ] \times \\
( \int_{0}^{\infty } exp(-\frac{3u^{2}}{2\overline{v}^{2}_{\chi}}) sinh(\frac{3uv_{\ast}}{\overline{v}^{2}_{\chi}}) \: \times \\
  \left \{ exp(-\frac{m_{\chi}u^{2}}{2E_{0}}) - exp(-\frac{m_{\chi}u^{2}}{2E_{0}}\frac{\mu_{i}}{\mu^{2}_{+,i}}) exp(-\frac{m_{\chi}v_{e}^{2}}{2E_{0}}\frac{\mu_{i}}{\mu^{2}_{-,i}}(1-\frac{\mu_{i}}{\mu^{2}_{+,i}})) \right \} \times \\
  du )
\end{multline}

Though it seems impossible to evaluate equations \ref{CR_by_hydrogen} and \ref{CR_by_hevier} analytically, but it is possible to evaluate them using the state-of-the-art stellar evolutionary codes. In this study, we used version 12778 of the MESA stellar evolutionary code to calculate CR by stars. MESA is a free and open-source stellar evolutionary code that can simulate stars from very low-mass ones to the very high-mass ones ($ \approx 10^{-3} - 10^{3} M_{\odot} $) . The full capabilities of MESA are documented in its official instrument papers \citep{2013ApJS..208....4P,MESA_2018,MESA_2015, MESA_2019,2011ApJS..192....3P} .

\subsection{Dark matter luminosity} \label{S_CR_by_hevier}
If DM particles annihilate inside stars then they can act as a new source of energy inside stars. By multiplying the CR relations by the $m_{\chi} \: c^{2}$ it is possible to calculate the luminosity that is produced by this way:

\begin{equation} \label{Equa_L_x}
L_{x} = CR \times \: m_{\chi} \: c^{2}
\end{equation}

\subsection{Dark matter density profile} \label{Sec_DM_density_profile}
\begin{figure*}
	\centering
	\includegraphics[width=1.4\columnwidth]{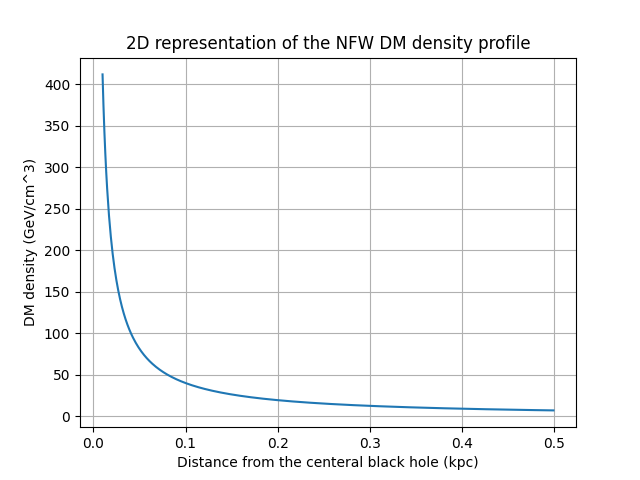}
	\caption{\label{Fig_NFW_DM_Profile} 2D representation of the NFW DM density profile for central regions of the Milky way galaxy. From the Figure it is conceivable that for regions far from the center, DM density reduces rapidly to near-zero. }
\end{figure*}

N-body simulations of galaxies in the standard model of cosmology (i.e. $ \Lambda CDM \: model $) reveal the non-uniform distribution of DM inside galaxies \citep{Merritt2006, Navarro1996}. NFW DM density profile gives the radial distribution of DM inside galaxies and is written in the form \citep{Lin2019} :
\begin{equation}  \label{eq_NFW_profile}
\rho_{NFW}(r) =  \frac{\rho_{0}}{\frac{r}{r_{0}}\left ( 1 + \frac{r}{r_{0}} \right )^{2}}
\end{equation}
In the case of milky way galaxy $\rho_{0} \: = 0.51 \: GeV \: cm^{-3}$ is the DM density around the sun and $r_{0} \: = 8.1 \: kpc$ is the distance of the sun from the central black hole of the Galaxy. Figure \ref{Fig_NFW_DM_Profile} depicts the 2D representation of the Equation \ref{eq_NFW_profile} for central regions of our galaxy. From the figure it is conceivable that for regions far from the center, DM density reduces rapidly to near-zero. This behaviour emphasises the importance of DM effects on stars that are evolving in regions near the Galactic massive black hole.

\subsection{Dynamic of binary star systems} \label{Dynamic_of_binary_star_systems}
According to the equations \ref{CR_by_hydrogen} and \ref{CR_by_hevier} , CR by stars is a function of the speed of the stars 
$v_{\ast}$ . Then, CR by each star within the BSS will change while stars orbit around each other in an elliptical motion. In this section, we review the necessary equations that are needed to describe the motion of stars in BSSs. \\
If two stars with masses $M_{1}$ and $M_{2}$ orbit around each other in an elliptical motion with semi-major axis a and ellipticity e, then the orbital period of the system can be evaluated using Kepler's third law \citep{hilditch_2001}:

\begin{equation}
P^{2} = \dfrac{4 \pi^{2} a^{3}}{G M}
\end{equation}

where $M \: (= M_{1} + M_{2})$ is the total mass of the system. Speed of stars in periastron and apastron can be calculated using \citep{hilditch_2001} (see Fig \ref{Fig_Apastron_Peristron}):

\begin{figure*}
	\centering
	\includegraphics[width=1.2\columnwidth]{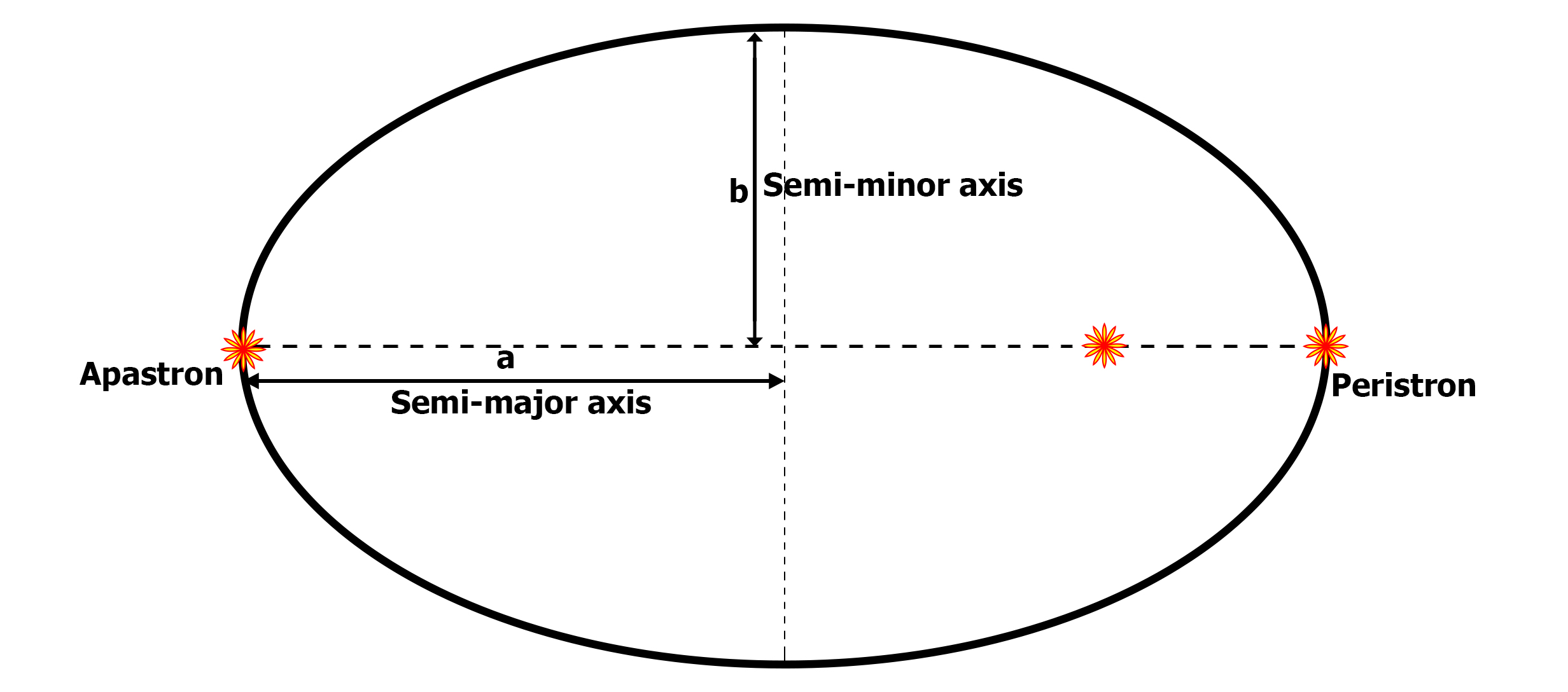}
	\caption{\label{Fig_Apastron_Peristron} Schematic view of a BSS and positions of periastron and apastron. }
\end{figure*}

\begin{equation} \label{v_p}
V_{p} = \sqrt{\frac{GM(1+e)}{a(1-e)}}
\end{equation}

and

\begin{equation} \label{v_a}
V_{a} = \sqrt{\frac{GM(1-e)}{a(1+e)}}
\end{equation}

\section{Effects of binary star parameters on the CR of DM particles} \label{binary_system_parameters}
In this section, we investigated BSSs parameter effects on the CR of DM particles. In binary systems, the speed of the stars is not constant, as they usually follow elliptical motion rather than circular. This speed variation causes the periodic changes in the CR by each star and also periodic changes in the total CR by the system. According to Equations \ref{CR_by_hydrogen} and \ref{CR_by_hevier} CR by each star is a complex function of the speed of the each star $v_{\ast}$. In fact, CR has the functionality of the form:
\begin{equation} \label{CR_heavier_functionality}
C_{\chi ,(i,H)} \propto \frac{1}{v_{\ast}} \times exp((...) \times -v^{2}_{\ast}) \times \int_{0}^{\infty } (...) \times sinh((...) \times v_{\ast}) du 
\end{equation}
for both hydrogen atoms and heavier elements. In addition, according to Equations \ref{v_p} and \ref{v_a}, in BSSs, the speed of each star is a function of different parameters of a BSSs. In fact $v_{\ast}$ has the functionality of the form $v_{\ast} \propto \sqrt{M}$ with total mass of a BSS, and $v_{\ast} \propto \sqrt{\frac{1}{a}}$ with semi-major axis and  $v_{\ast} \propto \sqrt{\frac{(1+e)}{(1-e)}}$ with eccentricity. As a result, it is not easy to discuss about a general behaviour of the CR as a function of the speed of stars or its dependence on parameters of BSSs. \\
In the coming sections, to study the effects of BSSs parameters on CR, we keep all parameters of a BSS to be constants except the parameter that we are going to investigate (e.g $M$, $a$, $e$ or $\rho_{\chi}$). During the research, we calculated CR by stars when they are in the zero-age main-sequence phase (ZAMS) \citep{Dotter_2016} . Also, we supposed that binary components have consisted of a 
combination of a low-mass star (1.0 $M_{\odot}$), an intermediate-mass star (5.0 $M_{\odot}$) and a high-mass star (50.0 
$M_{\odot}$). We supposed that DM composed of WIMP particles with masses 100 $ Gev \: 
c^{-2} $.

\subsection{Effect of stellar masses: M} \label{Effect_of_stellar_masses}
\begin{figure*}
	\centering
	\includegraphics[width=1.4\columnwidth]{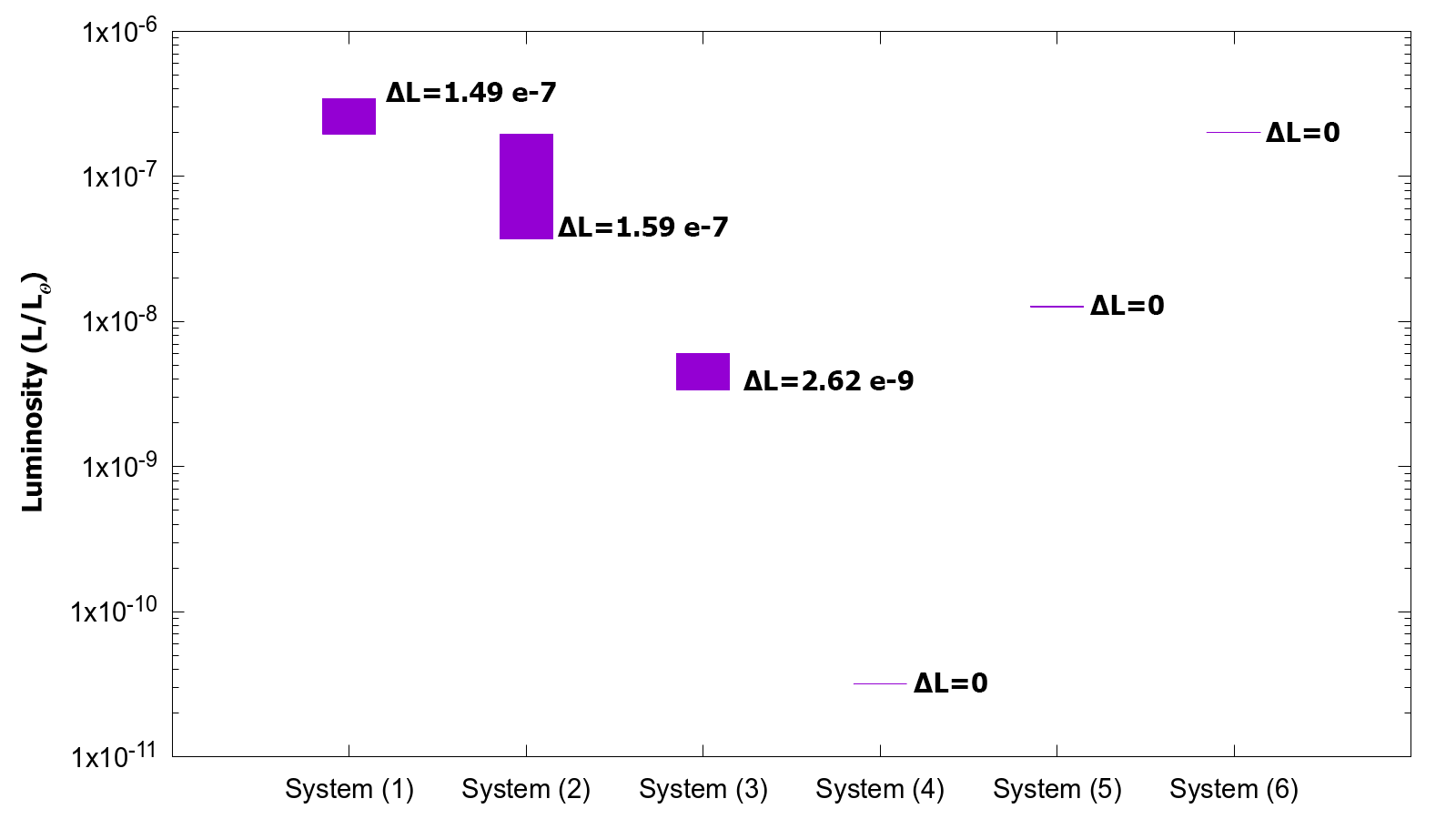}
	\caption{\label{Fig_L_table_1} Luminosity-variation of BSSs of Table \ref{table_effect_of_stellar_masses} that is produced by DM annihilation in BSSs. }
\end{figure*}

\begin{figure*}
	\centering
	\includegraphics[width=1.4\columnwidth]{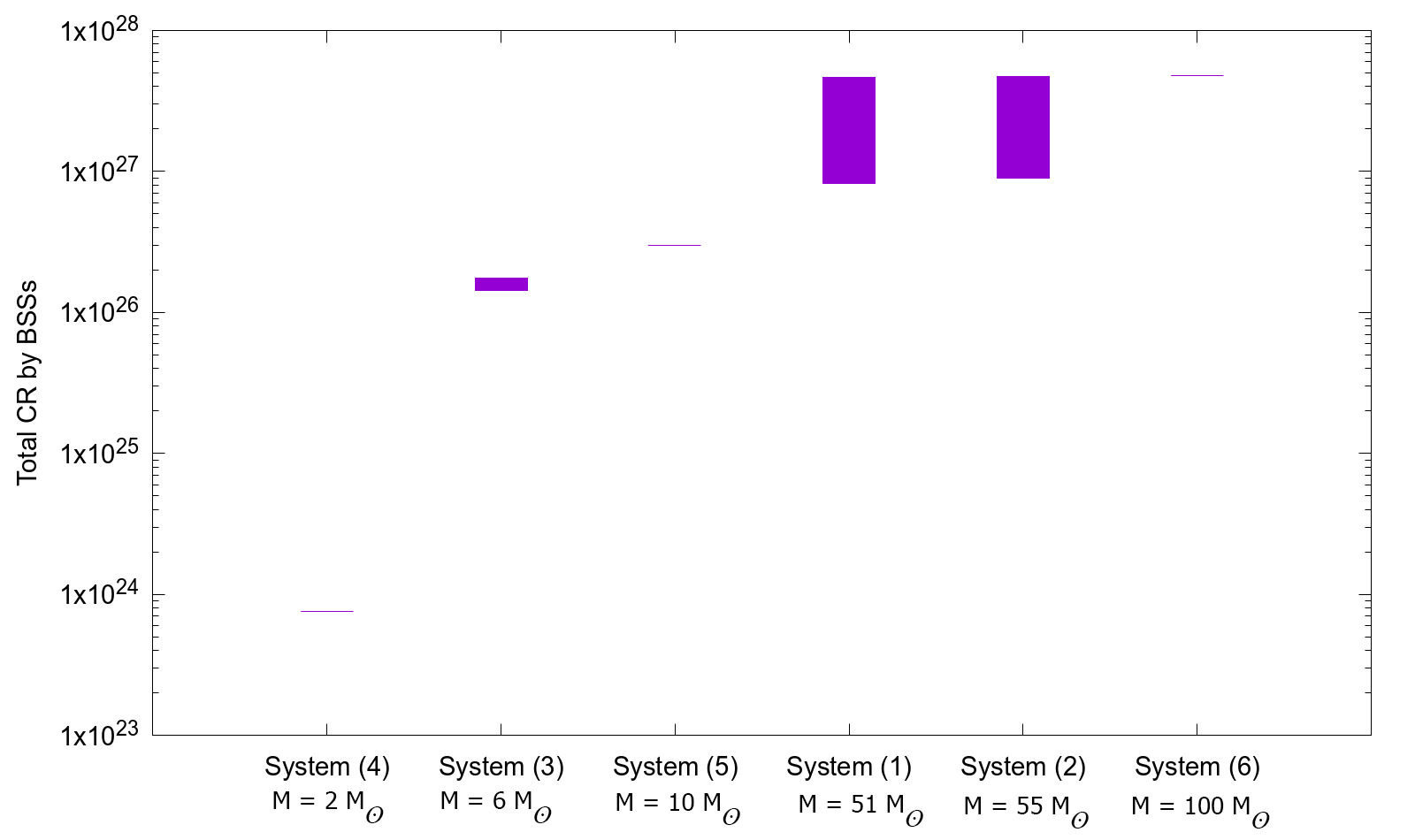}
	\caption{\label{Fig_L_table_1_b} Total CR by BSSs of Table \ref{table_effect_of_stellar_masses}. By increasing the total mass of a BSS (i.e. $M=M_{1}+M_{2}$), the total CR increases too.}
\end{figure*}

Using MESA stellar evolutionary code, CR by BSSs with different stellar masses are calculated and the results are summarized in Table \ref{table_effect_of_stellar_masses}  and Figures \ref{Fig_L_table_1} and \ref{Fig_L_table_1_b}. In Table \ref{table_effect_of_stellar_masses}, the eccentricity of all systems considered to be $ e = 0.9 $ and the semi-major axes considered to be $ a = 10 \: AU $. Density of DM around BSSs supposed to be   $\rho_{\chi} = 10^{3} \: Gev \: c^{-2}cm^{-3} $. The overall results are:

\begin{itemize}
  \item When stars are in apastron, they captures more DM particles in comparison to the time when they are in periastron (for instance, in Table \ref{table_effect_of_stellar_masses}, compare $T_{1}$ and $T_{4}$ which are capture rate for a $ 1 \: M_{\odot} $ star; or $T_{2}$ and $T_{3}$ which are capture rate for a $ 50 \: M_{\odot} $ star).
  
  \item Acoording to the total CR amounts that are presented in Table \ref{table_effect_of_stellar_masses}, the bigger the total mass of the system ($M = M_{1} + M_{2}$) is then the bigger the total CR is. This result  is demonstrated in graphic form in Figure \ref{Fig_L_table_1_b} too. As discussed above, this result is not general. Paying attention to Equation \ref{CR_heavier_functionality} and the fact that $v_{\ast} \propto \sqrt{M}$ , CR by a BSS is a more complex function of the total mass of the BSS.

  \item The results of the simulations show that, the most striking CR variation occurs for systems with the highest total mass ($M = M_{1} + M_{2}$). In System (4) (in Table \ref{table_effect_of_stellar_masses}), with the lowest total mass $M_{1} + M_{2} = 2.0 \: M_{\odot}$, the CR variation is:
\begin{align}
for \: 1.0 \: M_{\odot} \: star : \: \: \: \dfrac{T_{14}- T_{13}}{T_{13}} * 100 \simeq 7.12 \: \% .
\end{align}

For System (3) with increased total mass $M_{1} + M_{2} = 6.0 \: M_{\odot}$ , the CR variation increases to:
\begin{align}
for \: 1.0 \: M_{\odot} \: star:  \dfrac{T_{12}- T_{9}}{T_{9}} * 100 \simeq 22.96 \: \% \:\:\:\:\:\:\:   & \\
and \: for \: 5.0 M_{\odot} \: star:  \dfrac{T_{10}- T_{11}}{T_{10}} * 100 \simeq 23.08 \: \% .
\end{align} 

In System (5) with increased total mass $M_{1} + M_{2} = 10.0 \: M_{\odot}$ , the CR variation increases to:
\begin{align}
for \: 5.0 \: M_{\odot} \: star : \: \: \: \dfrac{T_{16}- T_{15}}{T_{15}} * 100 \simeq 40.8 \: \% .
\end{align} 

In System (1) with increased total mass $M_{1} + M_{2} = 51.0 \: M_{\odot}$ , the CR variation increases to:
\begin{align}
for \: 1.0 M_{\odot} \: star : \dfrac{T_{4}- T_{1}}{T_{1}} * 100 \simeq 484.96 \: \% \:\:\:\:\:\:\:   &  \\
and \: for \: 50.0 M_{\odot} \: star :  \dfrac{T_{2}- T_{3}}{T_{2}} * 100 \simeq 465.53 \: \% .
\end{align}  

In System (2) with increased total mass $M_{1} + M_{2} = 55.0 \: M_{\odot}$ the CR variation increases to:
\begin{align}
for \: 5.0 M_{\odot} \: star :  \dfrac{T_{8}- T_{5}}{T_{5}} * 100 \simeq 562.87 \: \% \:\:\:\:\:\:\:   &  \\
and \: for \: 50.0 M_{\odot} \: star:  \dfrac{T_{6}- T_{7}}{T_{6}} * 100 \simeq 548.12 \: \%.
\end{align}

And in System (6) with the highest total mass $M_{1} + M_{2} = 100.0 \: M_{\odot}$ the CR variation is the highest:
\begin{align}
for \: 50.0 \: M_{\odot} \: star : \: \: \: \dfrac{T_{18}- T_{17}}{T_{17}} * 100 \simeq 2893.55 \: \% .
\end{align}

\item Using Equation \ref{Equa_L_x} it is possible to calculate the luminosity-variation that is produced by DM annihilation in BSSs. The luminosity-variation for systems in Table \ref{table_effect_of_stellar_masses} are presented in Figure \ref{Fig_L_table_1}. For System (2), with the highest total mass ($M=M_{1} + M_{2}=55M_{\odot}$), the luminosity-variation is the highest, But for Systems (4)-(6) it is equal to zero. 

\end{itemize}

\begin{table*}
\centering
\caption{CR in BSSs with equal and unequal (last 3 rows) stellar-mass components. By increasing the total mass of a BSS ($M=M_{1}+M_{2}$) the total CR by the system increases too. Figure \ref{Fig_L_table_1_b} depicts a graphic representation of the amounts of this table too.}
\begin{threeparttable}
\begin{tabular}{|| c | c || c | c || c | c | c | c || }
\hline \hline
\shortstack{ $M_{1}$ \\ $(M_{\odot})$ } & \shortstack{ $M_{2}$ \\ $(M_{\odot})$ } & \shortstack{ $v_{1,p}$ \tnote{*} \\ $(m.sec^{-1})$ }  & \shortstack{ $v_{2,a}$ \tnote{**} \\ $(m.sec^{-1})$ } & \shortstack{ $CR \: by  \:M_{1}$ \tnote{***} \\ ($sec^{-1}) $ } & \shortstack{ $CR \: by \: M_{2}$ \tnote{****} \\ $(sec^{-1})$ } & \shortstack{ $CR_{M}$ \tnote{*****} \\ $(sec^{-1})$ } & \shortstack{system \\ number}  \\ \hline \hline
1.0  &  50.0  & 293558  & 15450  & $T_{1}= 6.65 \times 10^{22}$ & $T_{2}= 4.66 \times 10^{27}$ & $T_{1+2} = 4.66 \times 10^{27}$ & system  \\

50.0  &  1.0 & 293558  & 15450 & $T_{3}= 8.24 \times 10^{26}$     & $T_{4}= 3.89 \times 10^{23} $ & $T_{3+4} = 8.24 \times 10^{26}$ & (1) \\ \hline \hline

5.0  &  50.0  & 304853 & 16045 &  $T_{5}= 2.64 \times 10^{25}$     &    $T_{6}= 4.66  \times 10^{27}$  & $T_{5+6} = 4.69 \times 10^{27} $ & system \\

50.0  &  5.0  & 304853 & 16045 &  $T_{7}= 7.19 \times 10^{26}$     &    $T_{8}=  1.75 \times 10^{26}$  & $T_{7+8} = 8.94 \times 10^{26} $ & (2) \\ \hline \hline

1.0  &  5.0  & 100690 & 5299 &  $T_{9}= 3.18 \times 10^{23}$     &    $T_{10}=  1.76 \times 10^{26}$  & $T_{9+10} = 1.76 \times 10^{26} $ & system \\

5.0  &  1.0  & 100690 & 5299 &  $T_{11}= 1.43 \times 10^{26}$     &    $T_{12}= 3.91 \times 10^{23}$  & $T_{11+12} = 1.43 \times 10^{26} $ & (3) \\ \hline \hline

1.0  &  1.0  & 58133 & 3060 &  $T_{13}= 3.65 \times 10^{23}$     &    $T_{14}= 3.91 \times 10^{23}$  & $T_{13+14} = 7.56 \times 10^{23} $ & system (4) \\

5.0  &  5.0  & 129990 & 6842 &  $T_{15}= 1.25 \times 10^{26}$   &    $T_{16}= 1.76 \times 10^{26}$  & $T_{15+16} = 3.01 \times 10^{26} $ & system (5) \\

50.0  &  50.0  & 411064 & 21635  &  $T_{17}= 1.55 \times 10^{26}$    &    $T_{18}= 4.64 \times 10^{27}$  & $T_{17+18} = 4.79 \times 10^{27} $ & system (6) \\ \hline \hline
\end{tabular}
\begin{tablenotes}
\item[*] Speed of $ M_{1} $ star when it is in periastron.
\item[**] Speed of $ M_{2} $ star when it is in apastron.
\item[***] CR by $ M_{1} $ star when it is in periastron and when it is in ZAMS phase.
\item[****] CR by $ M_{2} $ star when it is in apastron and when it is in ZAMS phase.
\item[*****] Total CR : $ CR_{M} = CR_{M1} + CR_{M2} $
\end{tablenotes}
\end{threeparttable}
\label{table_effect_of_stellar_masses}
\end{table*}

\subsection{Effect of semi-major axis: a} \label{Effect_of_semi_major_axis}
\begin{figure*}
	\centering
	\includegraphics[width=1.4\columnwidth]{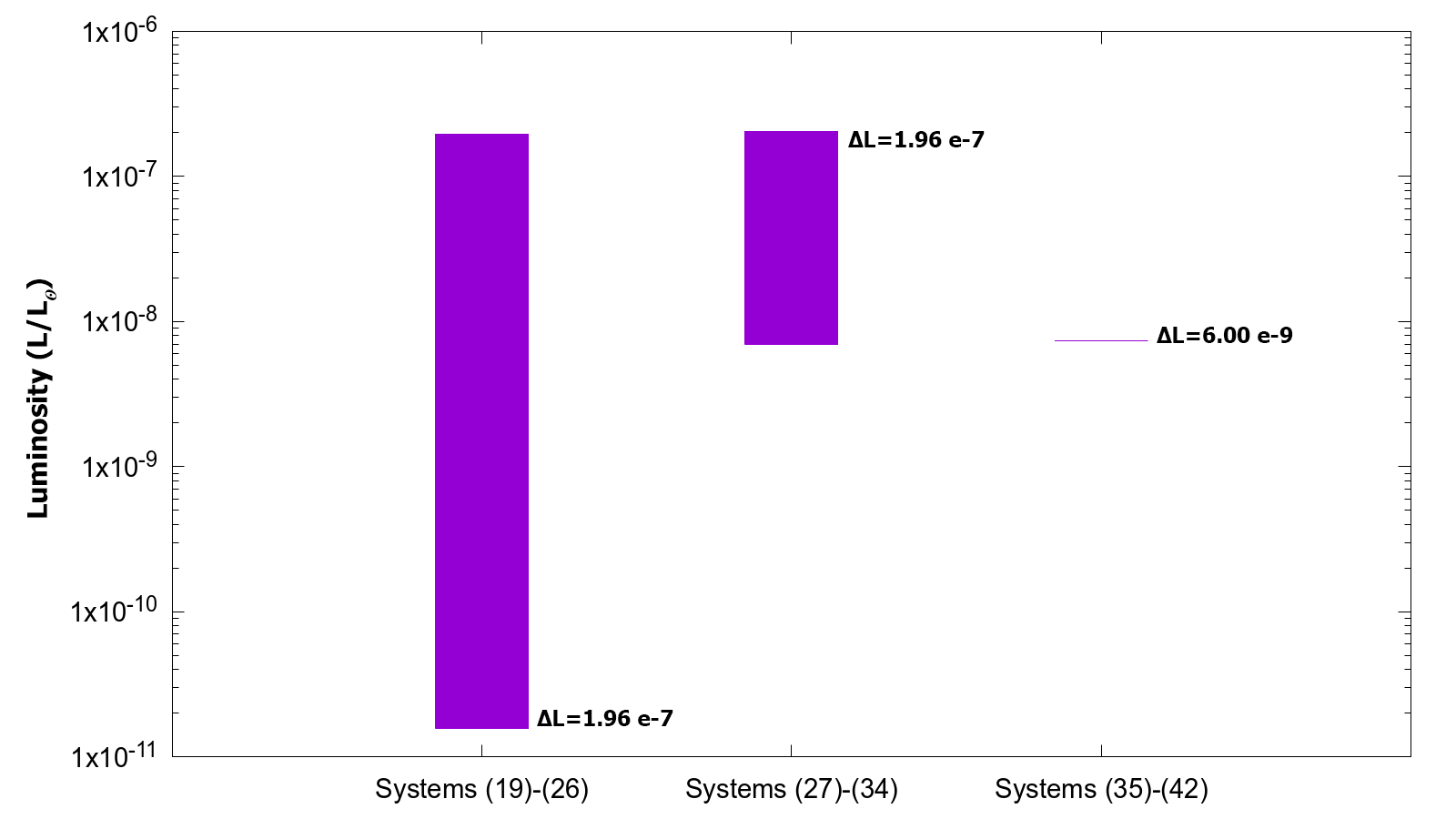}
	\caption{\label{Fig_L_table_3} Luminosity-variation of BSSs of Table \ref{table_effect_of_semi_major_axis_b} that is produced by DM annihilation in BSSs.}
\end{figure*}

In order to study the effects of semi-major axis "a" on the CR by BSSs, we keep all parameters of the systems to be constant, except the semi-major axis. The results of the simulations are presented in Table \ref{table_effect_of_semi_major_axis_a} for binaries with equal stellar-mass components and in Table  \ref{table_effect_of_semi_major_axis_b} for binaries with unequal stellar-mass components. In addition, using the total capture rate amounts $CR_M$ in Tables \ref{table_effect_of_semi_major_axis_a} and \ref{table_effect_of_semi_major_axis_b}, and also using Equation \ref{Equa_L_x}, the luminosity-variations for these systems are calculated and illustrated in Figures \ref{Fig_L_table_2} and \ref{Fig_L_table_3} , respectively. In all systems, the eccentricities of the systems considered to be constant (and equal to $ e = 0.9$)  and density of DM around BSSs supposed to be   $\rho_{\chi} = 10^{3} \: Gev \: c^{-2}cm^{-3} $. The overall results of the simulations are:

\begin{figure*}
	\centering
	\includegraphics[width=1.4\columnwidth]{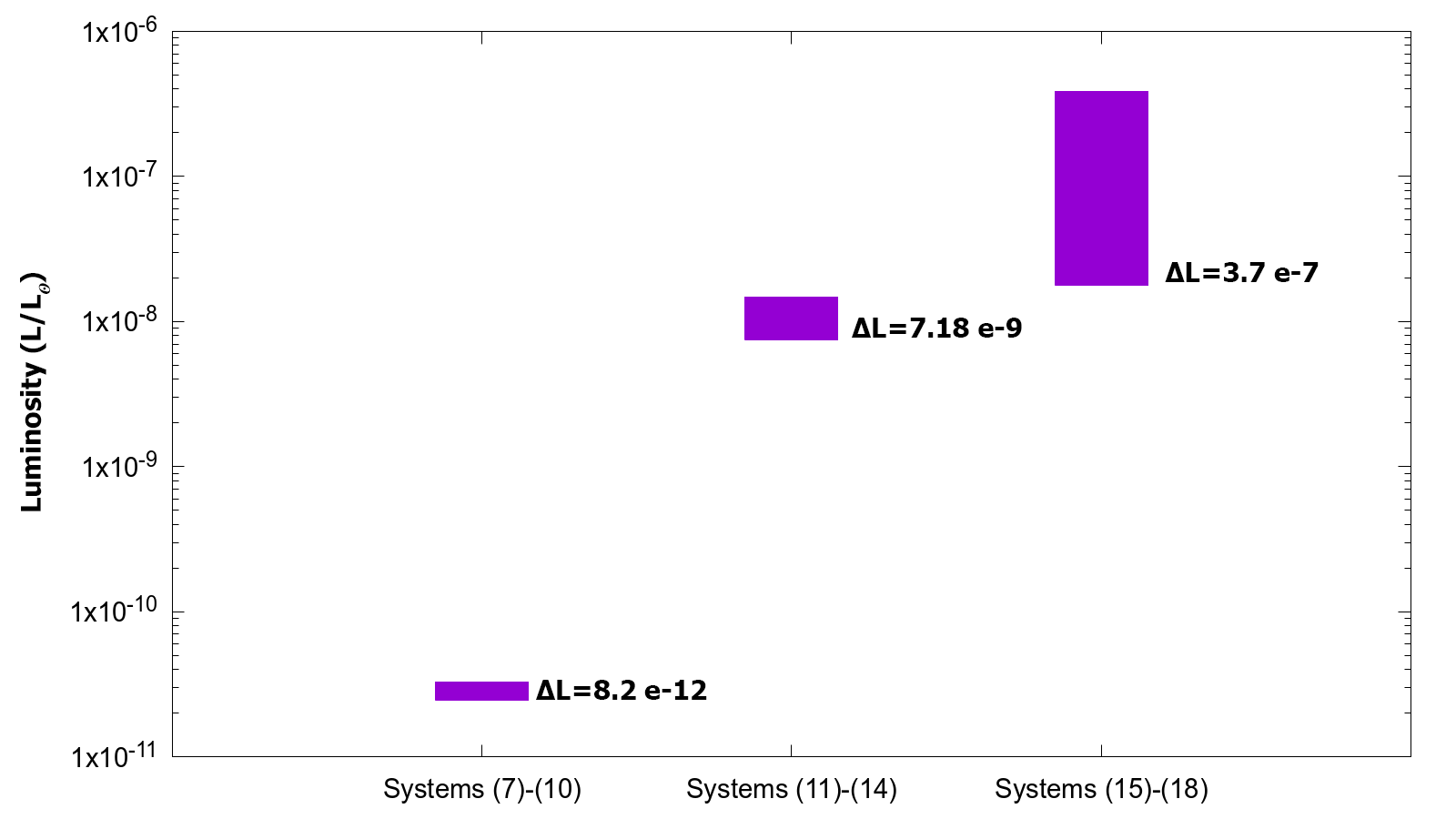}
	\caption{\label{Fig_L_table_2} Luminosity-variation of BSSs of Table \ref{table_effect_of_semi_major_axis_a} that is produced by DM annihilation in BSSs.}
\end{figure*}

\begin{figure*}
	\centering
	\includegraphics[width=1.4\columnwidth]{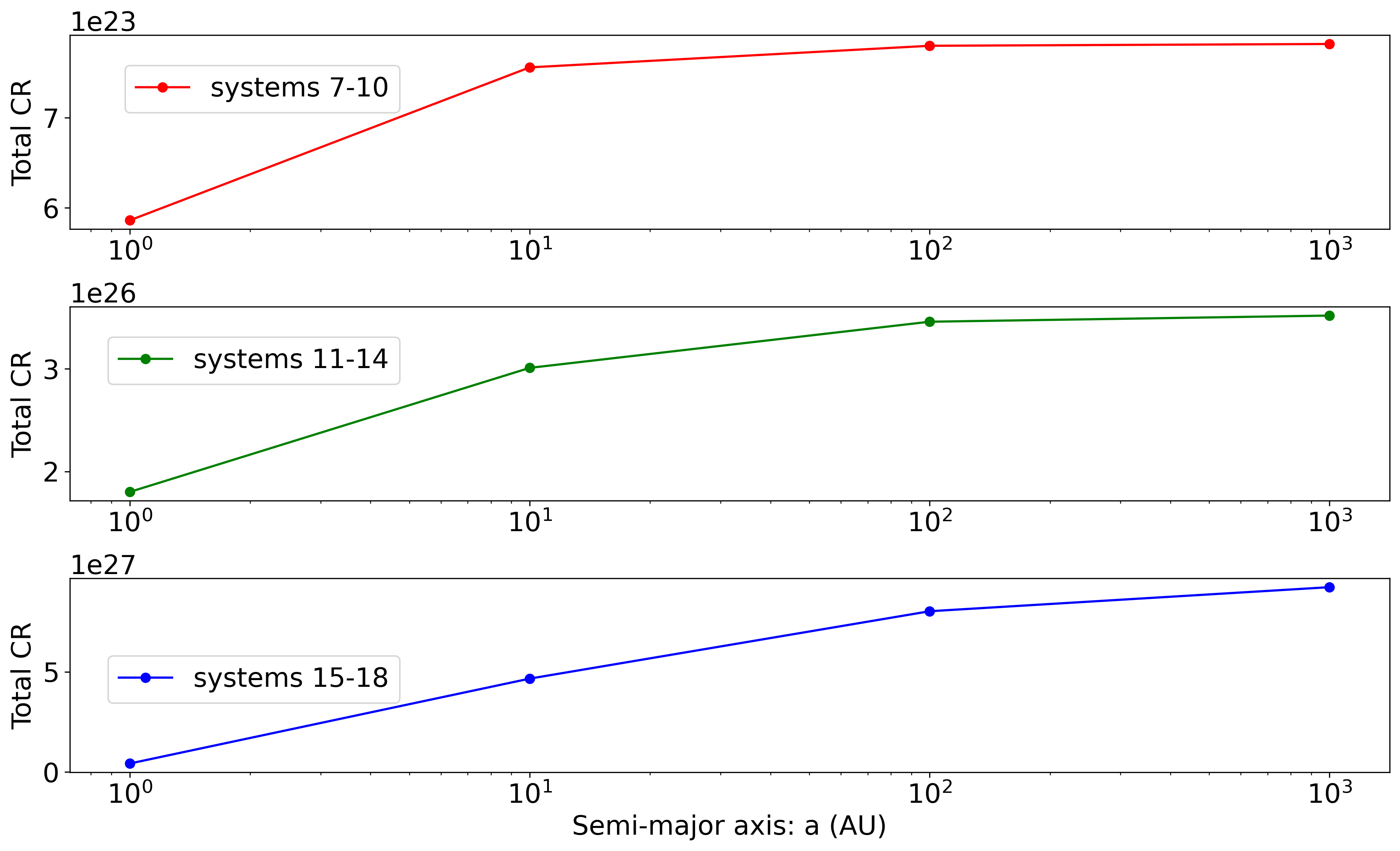}
	\caption{\label{Fig_table_2_b} Total CR by BSSs of Table \ref{table_effect_of_semi_major_axis_a}. From the figure it is conceivable that by increasing the semi-major axis of a BSS the total CR by the systems increases too.}
\end{figure*}

\begin{table*}
\centering
\caption{CR in BSSs with equal stellar-mass components and different semi-major axes. By increasing the semi-major axis of a BSS, the total CR by the system increases too. Figure \ref{Fig_table_2_b} depicts a graphic representation of the amounts of this table.}
\begin{threeparttable}
\begin{tabular}{|| c | c || c | c || c | c | c || }
\hline \hline
\shortstack{ $M_{1}$ \\ $(M_{\odot})$ } & \shortstack{ $M_{2}$ \\ $(M_{\odot})$ } & \shortstack{ $a$ \tnote{*} \\ $(AU)$ }  & \shortstack{ $ CR_{M1} $ \tnote{**} \\ $(sec^{-1})$ } & \shortstack{ $CR_{M2}$ \tnote{***} \\ ($sec^{-1}) $ } & \shortstack{ $CR_{M}$ \tnote{****} \\ $(sec^{-1})$ } &  \shortstack{system \\ number}  \\
\hline \hline
1.0  &  1.0  & 1 & $T_{19}= 1.95 \times 10^{23}$ & $T_{20}= 3.91 \times 10^{23}$ & $T_{19+20} = 5.86 \times 10^{23}$  & system (7) \\  
 & & 10 & $T_{21}= 3.65 \times 10^{23}$ & $T_{22}= 3.91 \times 10^{23}$ & $ T_{21+22} = 7.56 \times 10^{23} $ & system (8) \\  
 & & 100 & $T_{23}= 3.89 \times 10^{23}$ & $T_{24}= 3.91 \times 10^{23}$ & $ T_{23+24} = 7.80 \times 10^{23} $ & system (9) \\  
 & & 1000 & $T_{25}= 3.91 \times 10^{23}$ & $T_{26}= 3.91 \times 10^{23}$ & $ T_{25+26} = 7.82 \times 10^{23} $ & system (10) \\  \hline \hline

5.0 & 5.0 & 1 & $T_{27}= 5.59 \times 10^{24}$ & $T_{28}=  1.74 \times 10^{26}$ & $ T_{27+28} = 1.80 \times 10^{26} $ & system (11) \\ 
 & & 10 & $T_{29}= 1.25 \times 10^{26}$ & $T_{30}= 1.76 \times 10^{26}$ & $ T_{29+30} = 3.01 \times 10^{26} $ & system (12) \\ 
 & & 100 &  $T_{31}= 1.70 \times 10^{26}$     &    $T_{32}= 1.76 \times 10^{26}$  & $ T_{31+32} = 3.46 \times 10^{26} $ & system (13) \\ 
 & & 1000  &  $T_{33}= 1.75 \times 10^{26}$     &    $T_{34}= 1.76 \times 10^{26}$  & $ T_{33+34} = 3.52 \times 10^{26} $ & system (14) \\  \hline \hline

50.0  &  50.0  & 1  &  $T_{35}= -1.86 \times 10^{13}$     &    $T_{36}= 4.27 \times 10^{27}$  & $ T_{35+36} = 4.27 \times 10^{26} $ & system (15) \\ 
 & & 10 &  $T_{37}= 4.64 \times 10^{27}$     &    $T_{38}=  1.55 \times 10^{26}$  & $ T_{37+38} = 4.66 \times 10^{27} $ & system (16) \\ 
 & & 100 & $T_{39}= 3.33  \times 10^{27}$     &    $T_{40}=  4.68 \times 10^{27}$  & $ T_{39+40} = 8.02 \times 10^{27} $ & system (17) \\ 
 & & 1000 &  $T_{41}= 4.53 \times 10^{27}$     &    $T_{42}=  4.69 \times 10^{27}$  & $ T_{41+42} = 9.22 \times 10^{27} $ & system (18) \\  \hline \hline

\end{tabular}
\begin{tablenotes}
\item[*] Semi-major axis in astronomical unit (AU)
\item[**] CR by $ M_{1} $ star when it is in periastron and when it is in ZAMS phase.
\item[***] CR by $ M_{2} $ star when it is in apastron and when it is in ZAMS phase.
\item[****] Total CR : $ CR_{M} = CR_{M1} + CR_{M2} $
\end{tablenotes}
\end{threeparttable}
\label{table_effect_of_semi_major_axis_a}
\end{table*}

\begin{table*}
\centering
\caption{CR in BSSs with unequal stellar-mass components and different semi-major axes. By increasing the semi-major axis of a BSS the total CR by the system increases too. Figure \ref{Fig_table_3} depicts a graphic representation of the amounts of this table.}
\begin{threeparttable}
\begin{tabular}{|| c | c || c | c || c | c | c || }
\hline \hline
\shortstack{ $M_{1}$ \\ $(M_{\odot})$ } & \shortstack{ $M_{2}$ \\ $(M_{\odot})$ } & \shortstack{ $a$ \tnote{*} \\ $(AU)$ }  & \shortstack{ $ CR_{M1} $ \tnote{**} \\ $(sec^{-1})$ } & \shortstack{ $CR_{M2}$ \tnote{***} \\ ($sec^{-1}) $ } & \shortstack{ $CR_{M}$ \tnote{****} \\ $(sec^{-1})$ } &  \shortstack{system \\ number}  \\
\hline \hline
1.0  &  50.0  & 1 & $T_{43}= 7.8872 \times 10^{15}$ & $T_{44}= 4.4674 \times 10^{27}$ & $ T_{43+44} = 4.4674 \times 10^{27}$  & system (19) \\  
  &  & 10 &   $T_{45}= 6.6544 \times 10^{22}$ & $T_{46}= 4.6653 \times 10^{27}$ & $ T_{45+46} = 4.6654 \times 10^{27} $ & system (20) \\  
  &  & 100 &  $T_{47}= 3.2777 \times 10^{23}$ & $T_{48}= 4.6856 \times 10^{27}$ & $ T_{47+48} = 4.6859 \times 10^{27} $ & system (21) \\  
  &  & 1000 & $T_{49}= 3.8443 \times 10^{23}$ & $T_{50}= 4.6876 \times 10^{27}$ & $ T_{49+50} = 4.6880 \times 10^{27} $ & system (22) \\  \hline

50.0  &  1.0  & 1 & $T_{51}= 1.3027 \times 10^{20}$ & $T_{52}= 3.7256 \times 10^{23}$ & $ T_{51+52} = 3.7269 \times 10^{23} $ & system (23) \\ 
 & & 10 &     $T_{53}= 8.2445 \times 10^{26}$     & $T_{54}= 3.8938 \times 10^{23}$ & $ T_{53+54} = 8.2484 \times 10^{26} $ & system (24) \\ 
 & & 100 &    $T_{55}= 3.9398 \times 10^{27}$     &    $T_{56}= 3.9111 \times 10^{23}$  & $ T_{55+56} = 3.9402 \times 10^{27} $ & system (25) \\ 
 & & 1000  &  $T_{57}= 4.6071 \times 10^{27}$     &    $T_{58}= 3.9128 \times 10^{23}$  & $ T_{57+58} = 4.6075 \times 10^{27} $ & system (26) \\  \hline \hline \hline

5.0 & 50.0 & 1  &  $T_{59}= 8.7576 \times 10^{17}$     &    $T_{60}= 4.4506 \times 10^{27}$  & $ T_{59+60} = 4.4506 \times 10^{27} $ & system (27) \\ 
 & & 10 &    $T_{61}= 2.6406 \times 10^{25}$     &    $T_{62}= 4.6636 \times 10^{27}$  & $ T_{61+62} = 4.6900 \times 10^{27} $ & system (28) \\ 
 & & 100 &   $T_{63}= 1.4569 \times 10^{26}$    &    $T_{64}= 4.6854 \times 10^{27}$  & $ T_{63+64} = 4.8311 \times 10^{27} $ & system (29) \\ 
 & & 1000 &  $T_{65}= 1.7282 \times 10^{26}$     &    $T_{66}= 4.6876 \times 10^{27}$  & $ T_{65+66} = 4.8604 \times 10^{27} $ & system (30) \\  \hline

50.0 & 5.0 & 1  &  $T_{67}= 3.2578 \times 10^{19}$     &    $T_{68}= 1.6711 \times 10^{26}$  & $ T_{67+68} = 1.6711 \times 10^{26} $ & system (31) \\ 
 & & 10 &    $T_{69}= 7.1941 \times 10^{26}$     &    $T_{70}= 1.7520 \times 10^{26}$  & $ T_{69+70} = 8.9461 \times 10^{26} $ & system (32) \\ 
 & & 100 &   $T_{71}= 3.8865 \times 10^{27}$    &    $T_{72}= 1.7603 \times 10^{26}$  & $ T_{71+72} = 4.0625 \times 10^{27} $ & system (33) \\ 
 & & 1000 &  $T_{73}= 4.6008 \times 10^{27}$     &    $T_{74}= 1.7612 \times 10^{26}$  & $ T_{73+74} = 4.7769 \times 10^{27} $ & system (34) \\  \hline \hline \hline

1.0 & 5.0 & 1  &  $T_{75}= 4.8678 \times 10^{22}$     &    $T_{76}= 1.7512 \times 10^{26}$  & $ T_{75+76} = 1.7517 \times 10^{26} $ & system (35) \\ 
 & & 10 &    $T_{77}= 3.1768 \times 10^{23}$     &    $T_{78}= 1.7603 \times 10^{26}$  & $ T_{77+78} = 1.7635 \times 10^{26} $ & system (36) \\ 
 & & 100 &   $T_{79}= 3.8323 \times 10^{23}$    &    $T_{80}= 1.7612 \times 10^{26}$  & $ T_{79+80} = 1.7650 \times 10^{26} $ & system (37) \\ 
 & & 1000 &  $T_{81}= 3.9048 \times 10^{23}$     &    $T_{82}= 1.7613 \times 10^{26}$  & $ T_{81+82} = 1.7652 \times 10^{26} $ & system (38) \\  \hline

5.0 & 1.0 & 1  &  $T_{83}= 2.2221 \times 10^{25}$     &    $T_{84}= 3.8905 \times 10^{23}$  & $ T_{83+84} = 2.2610 \times 10^{25} $ & system (39) \\ 
 & & 10 &    $T_{85}= 1.4320 \times 10^{26}$     &    $T_{86}= 3.9107 \times 10^{23}$  & $ T_{85+86} = 1.4359 \times 10^{26} $ & system (40) \\ 
 & & 100 &   $T_{87}= 1.7252 \times 10^{26}$    &    $T_{88}= 3.9128 \times 10^{23}$  & $ T_{87+88} = 1.7291 \times 10^{26} $ & system (41) \\ 
 & & 1000 &  $T_{89}= 1.7576 \times 10^{26}$     &    $T_{90}= 3.9130 \times 10^{23}$  & $ T_{89+90} = 1.7615 \times 10^{26} $ & system (42) \\  \hline

\end{tabular}
\begin{tablenotes}
\item[*] Semi-major axis in astronomical unit (AU)
\item[**] CR by $ M_{1} $ star when it is in periastron and when it is in ZAMS phase.
\item[***] CR by $ M_{2} $ star when it is in apastron and when it is in ZAMS phase.
\item[****] Total CR : $ CR_{M} = CR_{M1} + CR_{M2} $
\end{tablenotes}
\end{threeparttable}
\label{table_effect_of_semi_major_axis_b}
\end{table*}

\begin{itemize}
\item CR by the $ 50 \: M_{\odot} $ star in the System (15) in Table \ref{table_effect_of_semi_major_axis_a} is negative, i.e. $ T_{35} = -1.8638 \times 10^{13} $. This means, the star losses DM particles instead of capturing them. The reason for being negative, in this case, is the very high speed of the $ 50 \: M_{\odot} $ star. Its speed at the periastron is $ v_{\ast} = 1299897 \: m \: sec^{-1} $ which is more than the escape velocity from the surface of the star: $ v_{\ast} > v_{esca} $ (When star is in ZAMS phase its escape velocity from its surface is $ v_{escape} = 273904 \: m \: sec^{-1} $. This speed is not presented in the Table \ref{table_effect_of_semi_major_axis_a}. We used MESA stellar evolutionary code to obtain the amount of $ v_{escape}$). As a result, we can say, in close BSSs, CR by stars can be negative. 

\item According to Figure \ref{Fig_L_table_2}, for BSSs with equal stellar-mass components, variation of semi-major axes will affect the luminosity of high-mass BSSs more than low-mass ones. A similar trend can be seen in Figure \ref{Fig_L_table_3} for binaries with unequal stellar-mass components. As a result, and in this case, massive BSSs are suitable cases to search for DM effects in BSSs.

\item According to the results of our simulations that are presented in Table \ref{table_effect_of_semi_major_axis_a} and Figure \ref{Fig_table_2_b} for systems with equal stellar-mass components and Table \ref{table_effect_of_semi_major_axis_b} and Figure \ref{Fig_table_3} for systems with unequal stellar-mass components, by increasing the semi-major axis of a BSS, the total CR by the system will increases too.

\begin{figure*}
	\centering
	\includegraphics[width=1.4\columnwidth]{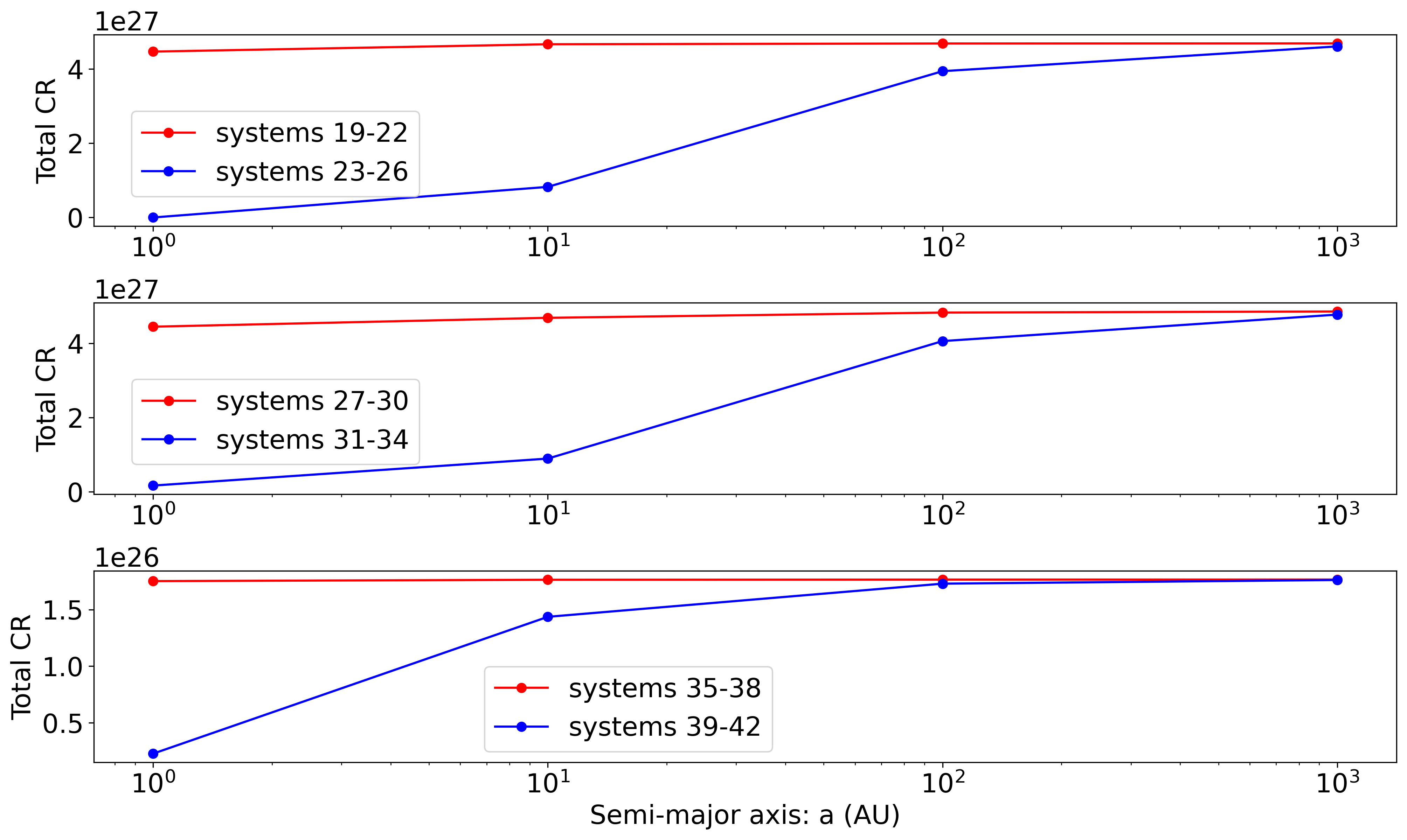}
	\caption{\label{Fig_table_3} Total CR by BSSs of the Table \ref{table_effect_of_semi_major_axis_b}. From the figure it is conceivable that by increasing the semi-major axes of a BSS the total CR by the systems increases too.}
\end{figure*}

\end{itemize}

\subsection{Effect of eccentricity: e} \label{Effect_of_eccentricity}
\begin{figure*}
	\centering
	\includegraphics[width=1.4\columnwidth]{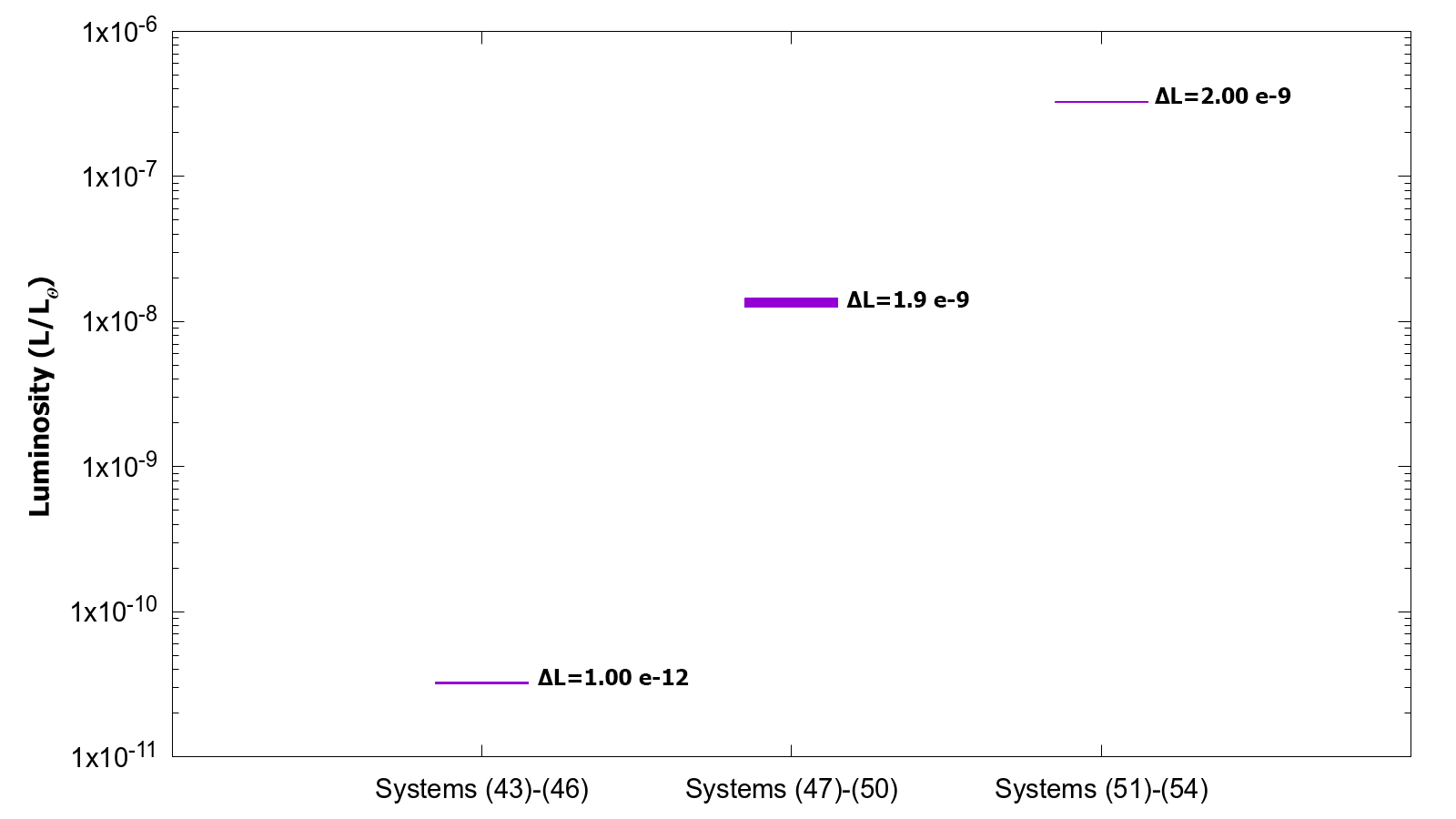}
	\caption{\label{Fig_L_table_4} Luminosity-variation of BSSs of Table \ref{table_effect_of_eccentricity_a} that is produced by DM annihilation in BSSs.}
\end{figure*}

\begin{figure*}
	\centering
	\includegraphics[width=1.4\columnwidth]{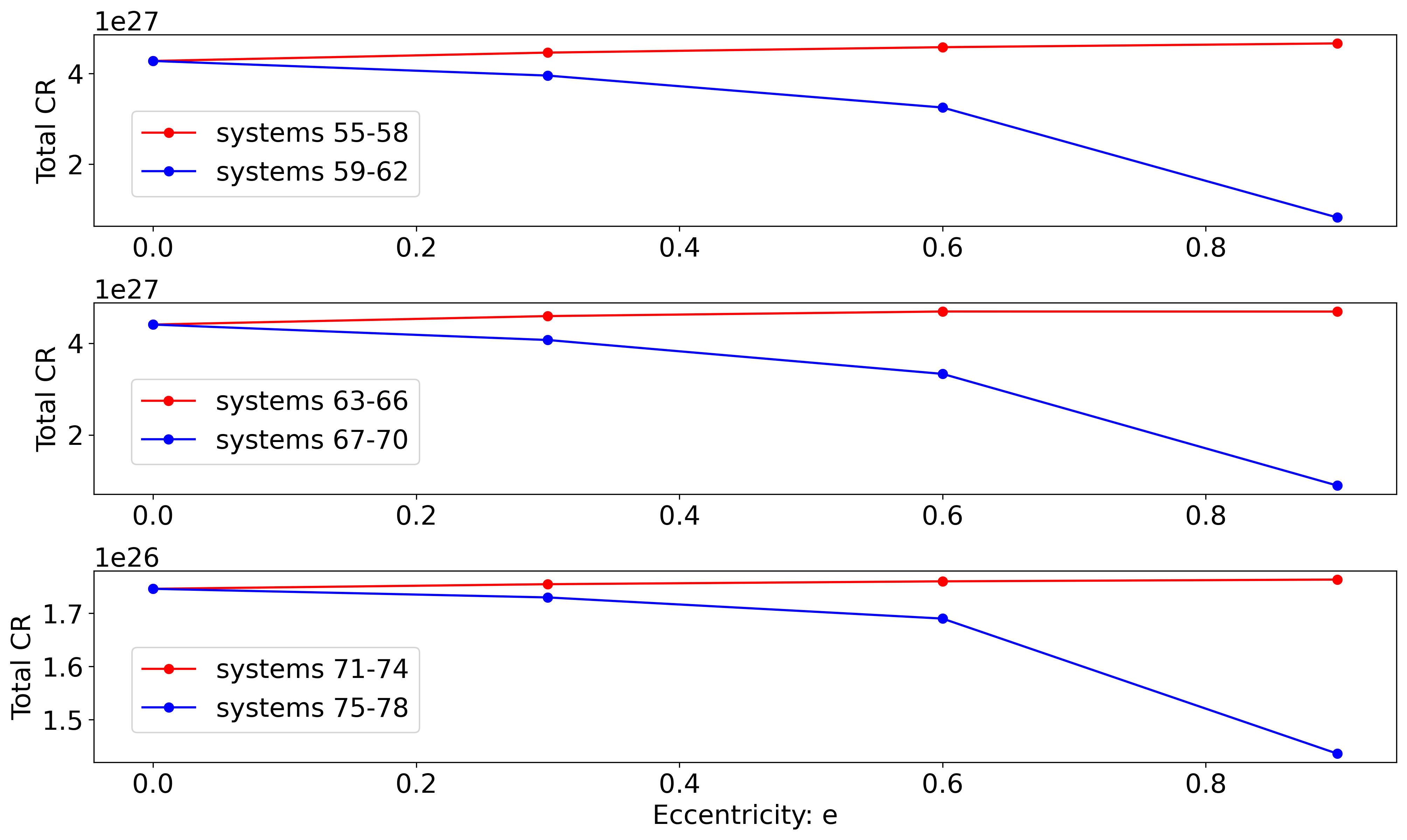}
	\caption{\label{Fig_5_b} Total CR by BSSs of the Table \ref{table_effect_of_eccentricity_b} (BSSs with unequall stellar-mass components).}
\end{figure*}

\begin{figure*}
	\centering
	\includegraphics[width=1.4\columnwidth]{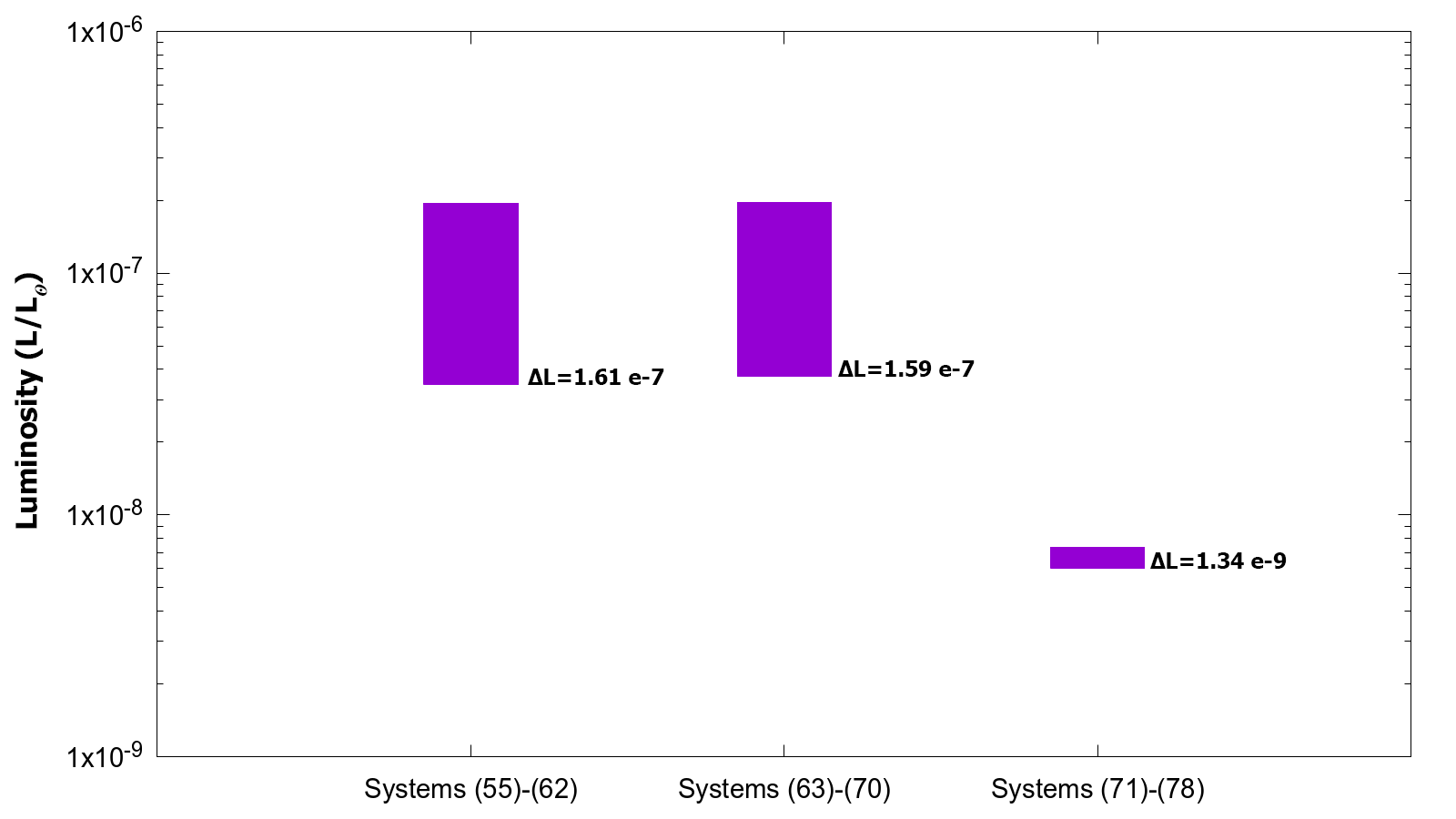}
	\caption{\label{Fig_L_table_5} Luminosity-variation of BSSs of Table \ref{table_effect_of_eccentricity_b} that is produced by DM annihilation in BSSs.}
\end{figure*}

In order to study the effects of eccentricity on the CR by BSSs, we keep all parameters of the BSSs to be constant, except the eccentricity. The results of the simulations are presented in Table \ref{table_effect_of_eccentricity_a} and Figure \ref{Fig_4_b} for binaries with equal stellar-mass components and in Table \ref{table_effect_of_eccentricity_b} and Figure \ref{Fig_5_b} for binaries with unequal stellar-mass components. In addition, using the total capture rate amounts $CR_M$ in Tables \ref{table_effect_of_eccentricity_a} and \ref{table_effect_of_eccentricity_b}, and also using Equation \ref{Equa_L_x}, the luminosity-variations for these systems are calculated and illustrated in Figures \ref{Fig_L_table_4} and \ref{Fig_L_table_5} , respectively. In all systems, the semi-major axes of the systems and DM density around the BSSs considered to be constant (i.e. $ a = 10 \: AU$ and $\rho_{\chi}=10^{3} \: Gev \: c^{-2} cm^{-3}$). The overall results of the simulations are:

\begin{figure*}
	\centering
	\includegraphics[width=1.4\columnwidth]{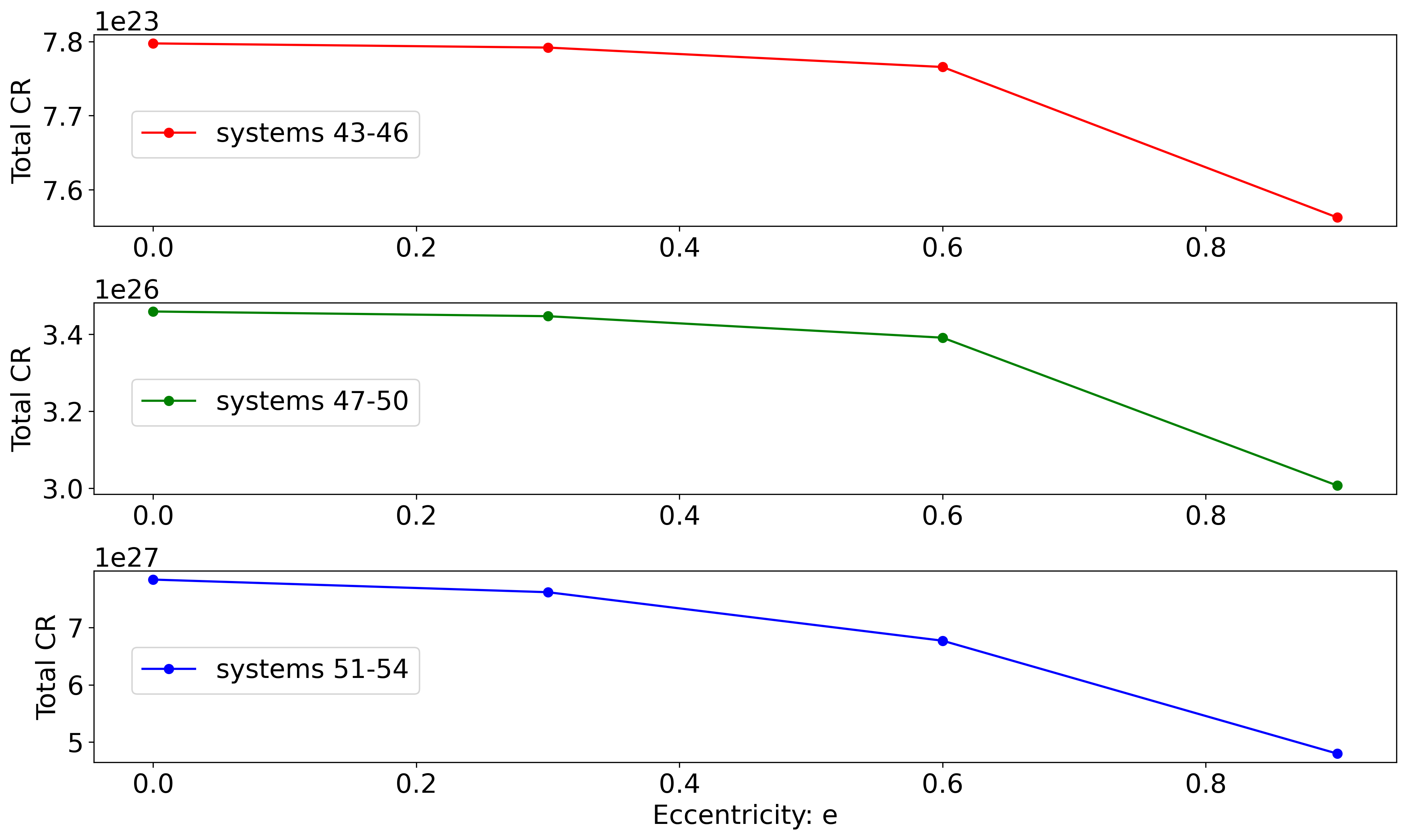}
	\caption{\label{Fig_4_b} Total CR by BSSs of the Table \ref{table_effect_of_eccentricity_a} (BSSs with equal stellar-mass components). By increasing the eccentricity of a system the total CR will decrease.}
\end{figure*}

\begin{table*}
\centering
\caption{CR in BSSs with equal stellar-mass components and different eccentricities. By increasing the eccentricity of a BSS the total CR by the system will decrease. Figure \ref{Fig_4_b} depicts a graphic representation of the amounts of this table.}
\begin{threeparttable}
\begin{tabular}{|| c | c || c | c || c | c | c || }
\hline \hline
\shortstack{ $M_{1}$ \\ $(M_{\odot})$ } & \shortstack{ $M_{2}$ \\ $(M_{\odot})$ } & \shortstack{ $e$ \tnote{*} }  & \shortstack{ $ CR_{M1} $ \tnote{**} \\ $(sec^{-1})$ } & \shortstack{ $CR_{M2}$ \tnote{***} \\ ($sec^{-1}) $ } & \shortstack{ $CR_{M}$ \tnote{****} \\ $(sec^{-1})$ } &  \shortstack{system \\ number}  \\
\hline \hline
1.0  &  1.0  & 0 (circle) & $T_{91}= 3.8987 \times 10^{23}$ & $T_{92}= 3.8987\times 10^{23}$ & $T_{91+92} = 7.7974 \times 10^{23}$  & system (43) \\  
 & & 0.3 & $T_{93}= 3.8865 \times 10^{23}$ & $T_{94}= 3.9053 \times 10^{23}$ & $ T_{93+94} = 7.7918 \times 10^{23} $ & system (44) \\  
 & & 0.6 & $T_{95}= 3.8562 \times 10^{23}$ & $T_{96}= 3.9094 \times 10^{23}$ & $ T_{95+96} = 7.7656 \times 10^{23} $ & system (45) \\  
 & & 0.9 & $T_{97}= 3.6504 \times 10^{23}$ & $T_{98}= 3.9122 \times 10^{23}$ & $ T_{97+98} = 7.5626 \times 10^{23} $ & system (46) \\  \hline \hline

5.0 & 5.0 & 0 (circle) & $T_{99}= 1.7296 \times 10^{26}$ & $T_{100}= 1.7296 \times 10^{26}$ & $ T_{99+100} = 3.4592 \times 10^{26} $ & system (47) \\ 
 & & 0.3 & $T_{101}= 1.7029 \times 10^{26}$ & $T_{102}= 1.7441 \times 10^{26}$ & $ T_{101+102} = 3.4470 \times 10^{26} $ & system (48) \\ 
 & & 0.6 &  $T_{103}= 1.6379 \times 10^{26}$     &    $T_{104}= 1.7533 \times 10^{26}$  & $ T_{103+104} = 3.3912 \times 10^{26} $ & system (49) \\ 
 & & 0.9  &  $T_{105}= 1.2474 \times 10^{26}$     &    $T_{106}= 1.7596 \times 10^{26}$  & $ T_{105+106} = 3.0070 \times 10^{26} $ & system (50) \\  \hline \hline

50.0  &  50.0  & 0 (circle)  &  $T_{107}= 3.9180 \times 10^{27}$     &    $T_{108}= 3.9180 \times 10^{27}$  & $ T_{107+108} = 7.8360 \times 10^{27} $ & system (51) \\ 
 & & 0.3 &  $T_{109}= 3.3596 \times 10^{27}$     &    $T_{110}= 4.2562 \times 10^{27}$  & $ T_{109+110} = 7.6158 \times 10^{27} $ & system (52) \\ 
 & & 0.6 & $T_{111}=  2.2874 \times 10^{27}$     &    $T_{112}= 4.4822  \times 10^{27}$  & $ T_{111+112} = 6.7696 \times 10^{27} $ & system (53) \\ 
 & & 0.9 &  $T_{113}= 1.5534 \times 10^{26}$     &    $T_{114}= 4.6438 \times 10^{27}$  & $ T_{113+114} = 4.7991 \times 10^{27} $ & system (54) \\  \hline \hline

\end{tabular}
\begin{tablenotes}
\item[*] Eccentricity.
\item[**] CR by $ M_{1} $ star when it is in periastron and when it is in ZAMS phase.
\item[***] CR by $ M_{2} $ star when it is in apastron and when it is in ZAMS phase.
\item[****] Total CR : $ CR_{M} = CR_{M1} + CR_{M2} $
\end{tablenotes}
\end{threeparttable}
\label{table_effect_of_eccentricity_a}
\end{table*}

\begin{table*}
\centering
\caption{CR in BSSs with unequal stellar-mass components and different eccentricities. Figure \ref{Fig_5_b} depicts a graphic representation of the amounts of this table.}
\begin{threeparttable}
\begin{tabular}{|| c | c || c | c || c | c | c || }
\hline \hline
\shortstack{ $M_{1}$ \\ $(M_{\odot})$ } & \shortstack{ $M_{2}$ \\ $(M_{\odot})$ } & \shortstack{ $e$ \tnote{*}  }  & \shortstack{ $ CR_{M1} $ \tnote{**} \\ $(sec^{-1})$ } & \shortstack{ $CR_{M2}$ \tnote{***} \\ ($sec^{-1}) $ } & \shortstack{ $CR_{M}$ \tnote{****} \\ $(sec^{-1})$ } &  \shortstack{system \\ number}  \\
\hline \hline
1.0  &  50.0  & 0 (circle) & $T_{115}= 3.5646 \times 10^{23}$ & $T_{116}= 4.2780 \times 10^{27}$ & $ T_{115+116} = 4.2784 \times 10^{27}$  & system (55) \\  
  &  & 0.3 &   $T_{117}= 3.2908 \times 10^{23}$ & $T_{118}= 4.4625 \times 10^{27}$ & $ T_{117+118} = 4.4628 \times 10^{27} $ & system (56) \\  
  &  & 0.6 &  $T_{119}= 2.6948 \times 10^{23}$ & $T_{120}= 4.5818 \times 10^{27}$ & $ T_{119+120} = 4.5821 \times 10^{27} $ & system (57) \\  
  &  & 0.9 & $T_{121}= 6.6544 \times 10^{22}$ & $T_{122}= 4.6653 \times 10^{27}$ & $ T_{121+122} = 4.6654 \times 10^{27} $ & system (58) \\  \hline

50.0  &  1.0  & 0 (circle) & $T_{123}= 4.2780 \times 10^{27}$ & $T_{124}= 3.5646 \times 10^{23}$ & $ T_{123+124} = 4.2784 \times 10^{27} $ & system (59) \\ 
 & & 0.3 &     $T_{125}= 3.9553 \times 10^{27}$     & $T_{126}= 3.7214 \times 10^{23}$ & $ T_{125+126} = 3.9557 \times 10^{27} $ & system (60) \\ 
 & & 0.6 &    $T_{127}= 3.2511 \times 10^{27}$     &    $T_{128}= 3.8228 \times 10^{23}$  & $ T_{127+128} = 3.2515 \times 10^{27} $ & system (61) \\ 
 & & 0.9  &  $T_{129}= 8.2445 \times 10^{26}$     &    $T_{130}= 3.8938 \times 10^{23}$  & $ T_{129+130} = 8.2484 \times 10^{26} $ & system (62) \\  \hline \hline \hline

5.0 & 50.0 & 0 (circle)  &  $T_{131}= 1.5939 \times 10^{26}$     &    $T_{132}= 4.2474 \times 10^{27}$  & $ T_{131+132} = 4.4068 \times 10^{27} $ & system (63) \\ 
 & & 0.3 &    $T_{133}= 1.4631 \times 10^{26}$     &    $T_{134}= 4.4453 \times 10^{27}$  & $ T_{133+134} = 4.5916 \times 10^{27} $ & system (64) \\ 
 & & 0.6 &   $T_{135}= 1.1813 \times 10^{26}$    &    $T_{136}= 4.5736 \times 10^{27}$  & $ T_{135+136} = 4.6917 \times 10^{27} $ & system (65) \\ 
 & & 0.9 &  $T_{137}= 2.6406 \times 10^{25}$     &    $T_{138}= 4.6636 \times 10^{27}$  & $ T_{137+138} = 4.6900 \times 10^{27} $ & system (66) \\  \hline

50.0 & 5.0 & 0 (circle) &  $T_{139}= 4.2474 \times 10^{27}$     &    $T_{140}= 1.5939 \times 10^{26}$  & $ T_{139+140} = 4.4068 \times 10^{27} $ & system (67) \\ 
 & & 0.3 &    $T_{141}= 3.9029 \times 10^{27}$     &    $T_{142}= 1.6691 \times 10^{26}$  & $ T_{141+142} = 4.0698 \times 10^{27} $ & system (68) \\ 
 & & 0.6 &   $T_{143}= 3.1592 \times 10^{27}$    &    $T_{144}= 1.7178 \times 10^{26}$  & $ T_{143+144} = 3.3310 \times 10^{27} $ & system (69) \\ 
 & & 0.9 &  $T_{145}= 7.1941 \times 10^{26}$     &    $T_{146}= 1.7520 \times 10^{26}$  & $ T_{145+146} = 8.9461 \times 10^{26} $ & system (70) \\  \hline \hline \hline

1.0 & 5.0 & 0 (circle) &  $T_{147}= 3.8703 \times 10^{23}$     &    $T_{148}= 1.7422 \times 10^{26}$  & $ T_{147+148} = 1.7461 \times 10^{26} $ & system (71) \\ 
 & & 0.3 &    $T_{149}= 3.8341 \times 10^{23}$     &    $T_{150}= 1.7510 \times 10^{26}$  & $ T_{149+150} = 1.7548 \times 10^{26} $ & system (72) \\ 
 & & 0.6 &   $T_{151}= 3.7450 \times 10^{23}$    &    $T_{152}= 1.7565 \times 10^{26}$  & $ T_{151+152} = 1.7602 \times 10^{26} $ & system (73) \\ 
 & & 0.9 &  $T_{153}= 3.1768 \times 10^{23}$     &    $T_{154}= 1.7603 \times 10^{26}$  & $ T_{153+154} = 1.7635 \times 10^{26} $ & system (74) \\  \hline

5.0 & 1.0 & 0 (circle) &  $T_{155}= 1.7422 \times 10^{26}$     &    $T_{156}= 3.8703 \times 10^{23}$  & $ T_{155+156} = 1.7461 \times 10^{26} $ & system (75) \\ 
 & & 0.3 &    $T_{157}= 1.7260 \times 10^{26}$     &    $T_{158}= 3.8899 \times 10^{23}$  & $ T_{157+158} = 1.7299 \times 10^{26} $ & system (76) \\ 
 & & 0.6 &   $T_{159}= 1.6862 \times 10^{26}$    &    $T_{160}= 3.9023 \times 10^{23}$  & $ T_{159+160} = 1.6901 \times 10^{26} $ & system (77) \\ 
 & & 0.9 &  $T_{161}= 1.4320 \times 10^{26}$     &    $T_{162}= 3.9107 \times 10^{23}$  & $ T_{161+162} = 1.4359 \times 10^{26} $ & system (78) \\  \hline

\end{tabular}
\begin{tablenotes}
\item[*] Eccentricity.
\item[**] CR by $ M_{1} $ star when it is in periastron and when it is in ZAMS phase.
\item[***] CR by $ M_{2} $ star when it is in apastron and when it is in ZAMS phase.
\item[****] Total CR : $ CR_{M} = CR_{M1} + CR_{M2} $
\end{tablenotes}
\end{threeparttable}
\label{table_effect_of_eccentricity_b}
\end{table*}

\begin{itemize}
\item According to the total CR amounts that are presented in Table \ref{table_effect_of_eccentricity_a} and Figure \ref{Fig_4_b}, in BSSs with equal stellar-mass components, by increasing the eccentricity of a system, the total CR decreases. This result is not correct for binaries with unequal stellar-mass components, as the results that are presented in Table \ref{table_effect_of_eccentricity_b} and Figure \ref{Fig_5_b} confirm this.

\item According to the CR amounts in Tables \ref{table_effect_of_eccentricity_a} and \ref{table_effect_of_eccentricity_b}, the most dramatic CR variations happens in the binaries with the highest eccentricities. For instance, in Table \ref{table_effect_of_eccentricity_a}, for $ 1.0\:M_{\odot}-1.0\:M_{\odot} $ BSS, the CR variations for different eccentricity configurations are (for systems (43)-(46)):
\begin{align}
for \: system \: (43), (e=0) \: \: \: : \: \: \: T_{92}- T_{91} = 0
\end{align}
\begin{align}
for \: system \: (44), (e=0.3) \: \: \: : \: \: \: T_{94}- T_{93} = 1.88 \times 10^{21}
\end{align}
\begin{align}
for \: system \: (45), (e=0.6) \: \: \: : \: \: \: T_{96}- T_{95} = 5.32 \times 10^{21}
\end{align}
\begin{align}
for \: system \: (46), (e=0.9) \: \: \: : \: \: \: T_{98}- T_{97} =  2.618 \times 10^{22}  .
\end{align}
The similar trend happens to $ 5.0\:M_{\odot}-5.0\:M_{\odot} $ and $ 50.0\:M_{\odot}-50.0\:M_{\odot} $ BSSs too. As an example, for binaries with unequal stellar-mass components, consider the $ 1.0\:M_{\odot}-50.0\:M_{\odot} $ system in Table \ref{table_effect_of_eccentricity_b} (Systems (55)-(58)). By increasing eccentricity in these systems CR variation in these systems will increase too:
\begin{align}
for \: system \: (55) \: \: \: : \: \: \: T_{116}- T_{115} = 4.2777 \times 10^{27}
\end{align}
\begin{align}
for \: system \: (56) \: \: \: : \: \: \: T_{118}- T_{117} = 4.4622 \times 10^{27}
\end{align}
\begin{align}
for \: system \: (57) \: \: \: : \: \: \: T_{120}- T_{119} = 4.5815 \times 10^{27}
\end{align}
\begin{align}
for \: system \: (58) \: \: \: : \: \: \: T_{122}- T_{121} = 4.6646 \times 10^{27} .
\end{align}
The similar trend happens to other systems in Table \ref{table_effect_of_eccentricity_b} too. As an important result of this section, the CR variation boosted when stars follow elliptical rather than circular orbits and this is true for all systems. Then looking for DM effects in BSSs with higher eccentricities are easier than looking for in lower eccentricity ones.

\item Figure \ref{Fig_L_table_4} demonstrates the luminosity-variations of the systems in the Table \ref{table_effect_of_eccentricity_a}. According to this figure, the luminosity-variation for high-mass systems is the highest. But this result is not true for the luminosity-variation of the systems in Table \ref{table_effect_of_eccentricity_b} (see Figure \ref{Fig_L_table_5} for the luminosity-variations of the systems of the Table \ref{table_effect_of_eccentricity_b}). This is because, according to Equations \ref{v_p} - \ref{v_a} (and using Equation \ref{Equa_L_x} too), luminosity-variation is not just a function of the total mass of the BSSs "$M$". But it is a function of the eccentricity "e" of the systems too.
\end{itemize}

\subsection{Effect of DM density: $\rho{\chi}$} \label{Effect_of_DM_density}
\begin{figure*}
	\centering
	\includegraphics[width=1.4\columnwidth]{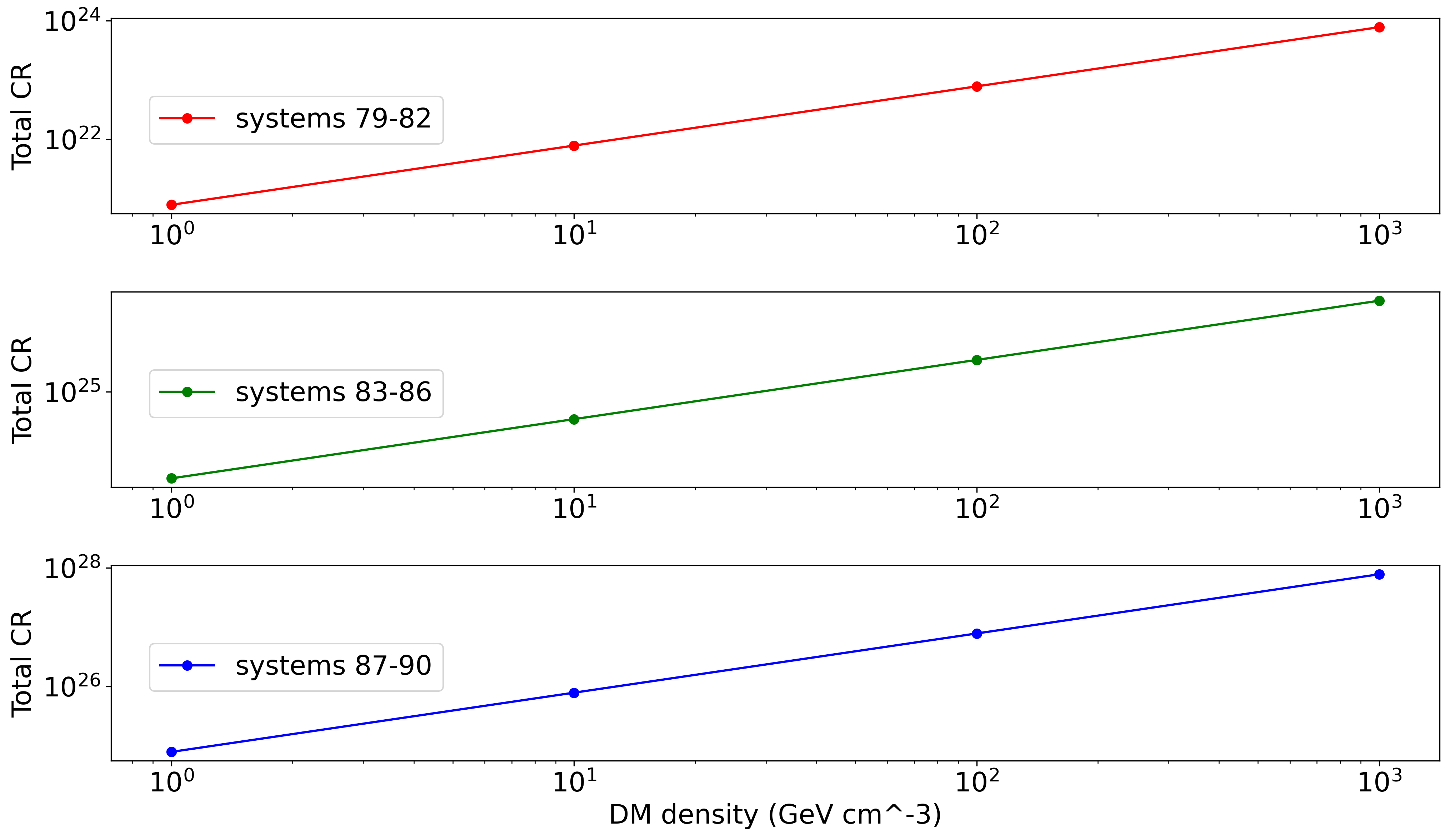}
	\caption{\label{fig_rho_table_6} Total CR by BSSs of the Table \ref{table_effect_of_DM_density_a} (BSSs with equal stellar-mass components). By increasing DM density around a BSS the total CR will increases too.}
\end{figure*}

\begin{figure*}
	\centering
	\includegraphics[width=1.4\columnwidth]{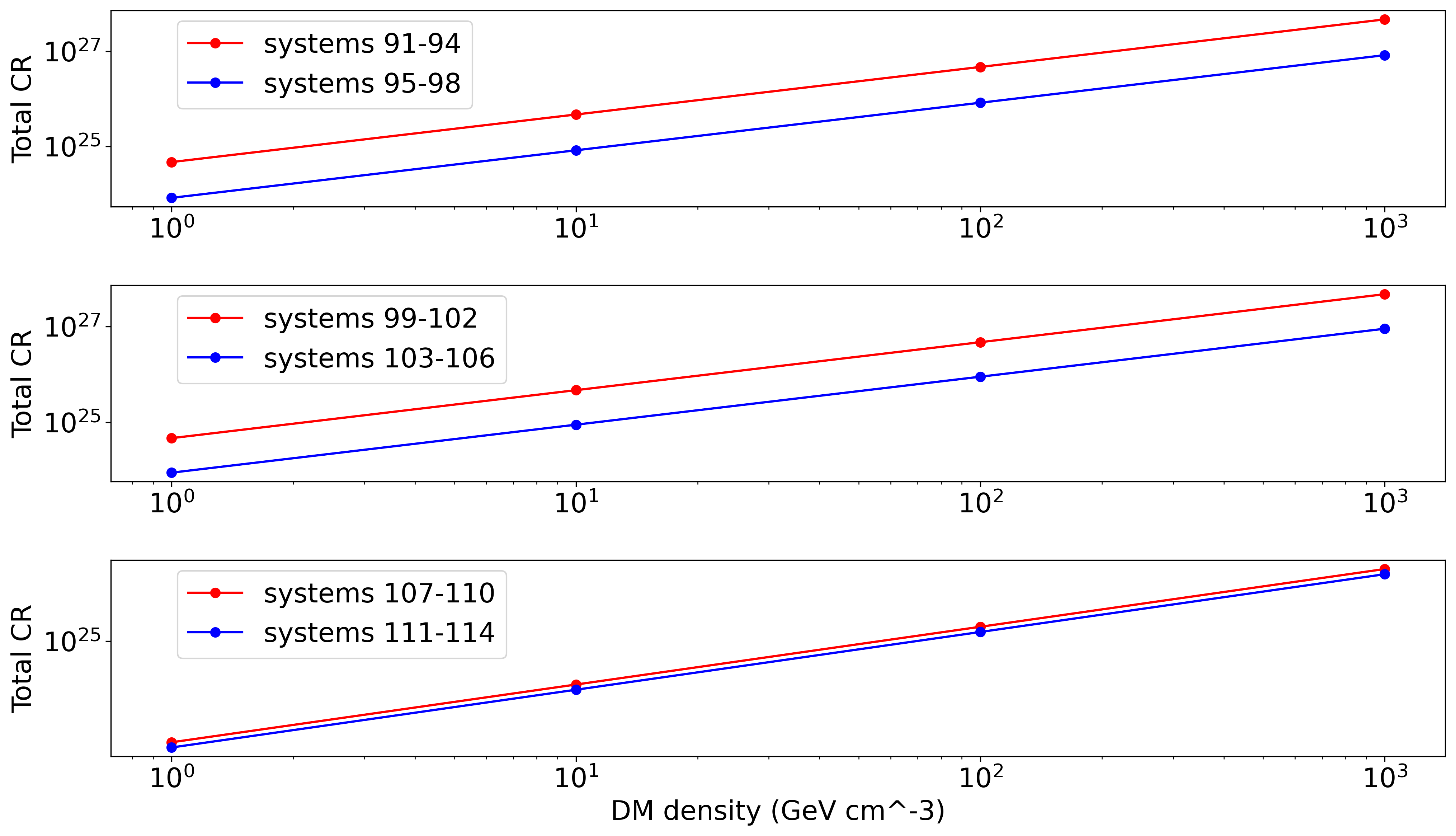}
	\caption{\label{fig_rho_table_7} Total CR by BSSs of the Table \ref{table_effect_of_DM_density_b} (BSSs with unequal stellar-mass components). By increasing DM density around a BSS the total CR will increases too.}
\end{figure*}

To study effects of DM density on the CR by BSSs, we keep all parameters of the systems to be constant, except DM density, $\rho{\chi}$ (i.e. we took eccentricity to be $e=0.9$ and semi-major axis to be $a=10 \: AU$ for all systems of this section). Using Equation \ref{eq_NFW_profile} one can calculate DM density in different locations of the Milky way galaxy. The results of the simulations are presented in Table \ref{table_effect_of_DM_density_a} and Figure \ref{fig_rho_table_6} for binaries with equal stellar-mass components and in Table \ref{table_effect_of_DM_density_b} and
Figure \ref{fig_rho_table_7} for binaries with unequal stellar-mass components. The overall results of the simulations are:
\begin{itemize}
\item According to the total CR amounts that are presented in Table \ref{table_effect_of_DM_density_a} and Figure \ref{fig_rho_table_6}, in BSSs with equall stellar-mass components, by increasing DM density around a BSS, the total CR by the systems increases too. The same behaviour can be seen in Table \ref{table_effect_of_DM_density_b} and Figure \ref{fig_rho_table_7} for systems with unequall stellar-mass components too. As can be seen in Figures \ref{fig_rho_table_6} and \ref{fig_rho_table_7} the increasion of total CR by the systems has linear relation with DM density. This behaviour can be conceivable by payying attention to the Equations \ref{CR_by_hydrogen} and \ref{CR_by_hevier}. According to these equations CR by stars have linear relation with DM density that sorrounds them:
\begin{align}
C_{\chi ,(H, i)} \: \propto \: \rho_{\chi}
\end{align}

\end{itemize}

\begin{table*}
\centering
\caption{CR in BSSs with equal stellar-mass components and different DM densities. By increasing the density of DM around a BSS the total CR by the system increases too. Figure \ref{fig_rho_table_6} depicts a graphic representation of the amounts of this table.}
\begin{threeparttable}
\begin{tabular}{|| c | c || c | c || c | c | c || }
\hline \hline
\shortstack{ $M_{1}$ \\ $(M_{\odot})$ } & \shortstack{ $M_{2}$ \\ $(M_{\odot})$ } & \shortstack{ $ \rho_{\chi} $ \tnote{*} \\ $(Gev \: c^{-2} \: cm^{-3})$ }  & \shortstack{ $ CR_{M1} $ \tnote{**} \\ $(sec^{-1})$ } & \shortstack{ $CR_{M2}$ \tnote{***} \\ ($sec^{-1}) $ } & \shortstack{ $CR_{M}$ \tnote{****} \\ $(sec^{-1})$ } &  \shortstack{system \\ number}  \\
\hline \hline
1.0  &  1.0  & 1000 & $T_{163}= 3.6504 \times 10^{23}$ & $T_{164}= 3.9122 \times 10^{23}$ & $T_{163+164} = 7.5626 \times 10^{23}$  & system (79) \\  
 & & 100                 & $T_{165}= 3.6504\times 10^{22}$  & $T_{166}= 3.9122 \times 10^{22}$ & $ T_{165+166} = 7.5626 \times 10^{22} $ & system (80) \\  
 & & 10                   & $T_{167}= 3.6504 \times 10^{21}$ & $T_{168}= 3.9122 \times 10^{21}$ & $ T_{167+168} = 7.5626 \times 10^{21} $ & system (81) \\  
 & & 1                     & $T_{169}= 3.6504 \times 10^{20}$ & $T_{170}= 3.9122 \times 10^{20}$ & $ T_{169+170} = 7.5626 \times 10^{20} $ & system (82) \\  \hline \hline

5.0 & 5.0 & 1000 & $T_{171}= 1.2474 \times 10^{26}$  & $T_{172}= 1.7596 \times 10^{26}$     & $ T_{171+172} = 3.0070 \times 10^{26} $ & system (83) \\ 
 & & 100              & $T_{173}= 1.2474 \times 10^{25}$  &    $T_{174}= 1.7596 \times 10^{25}$  & $ T_{173+174} = 3.0070 \times 10^{25} $ & system (84) \\ 
 & & 10                &  $T_{175}= 1.2474 \times 10^{24}$ &    $T_{176}= 1.7596 \times 10^{24}$  & $ T_{175+176} = 3.0070 \times 10^{24} $ & system (85) \\ 
 & & 1                  &  $T_{177}= 1.2474\times 10^{23}$  &    $T_{178}= 1.7596 \times 10^{23}$  & $ T_{177+178} = 3.0070 \times 10^{23} $ & system (86) \\  \hline \hline

50.0  &  50.0  & 1000  &  $T_{179}= 1.5534 \times 10^{26}$  &    $T_{180}= 4.6438 \times 10^{27}$   & $ T_{179+180} = 4.7991 \times 10^{27} $ & system (87) \\ 
 & & 100                      &  $T_{181}= 1.5534 \times 10^{25}$  &    $T_{182}= 4.6438\times 10^{26}$    & $ T_{181+182} = 4.7991 \times 10^{26} $ & system (88) \\ 
 & & 10                        & $T_{183}=  1.5534 \times 10^{24}$  &    $T_{184}= 4.6438  \times 10^{25}$  & $ T_{183+184} = 4.7991 \times 10^{25} $ & system (89) \\ 
 & & 1                          &  $T_{185}= 1.5534 \times 10^{23}$  &    $T_{186}= 4.6438\times 10^{24}$    & $ T_{185+186} = 4.7991 \times 10^{24} $ & system (90) \\  \hline \hline

\end{tabular}
\begin{tablenotes}
\item[*] DM density that sorround BSSs.
\item[**] CR by $ M_{1} $ star when it is in periastron and when it is in ZAMS phase.
\item[***] CR by $ M_{2} $ star when it is in apastron and when it is in ZAMS phase.
\item[****] Total CR : $ CR_{M} = CR_{M1} + CR_{M2} $
\end{tablenotes}
\end{threeparttable}
\label{table_effect_of_DM_density_a}
\end{table*}

\begin{table*}
\centering
\caption{CR in BSSs with unequal stellar-mass components and different DM densities. By increasing the density of DM around a BSS the total CR by the system increases too. Figure \ref{fig_rho_table_7} depicts a graphic representation of the amounts of this table.}
\begin{threeparttable}
\begin{tabular}{|| c | c || c | c || c | c | c || }
\hline \hline
\shortstack{ $M_{1}$ \\ $(M_{\odot})$ } & \shortstack{ $M_{2}$ \\ $(M_{\odot})$ } & \shortstack{ $ \rho_{\chi} $ \tnote{*} \\ $(Gev \: c^{-2} \: cm^{-3})$ }  & \shortstack{ $ CR_{M1} $ \tnote{**} \\ $(sec^{-1})$ } & \shortstack{ $CR_{M2}$ \tnote{***} \\ ($sec^{-1}) $ } & \shortstack{ $CR_{M}$ \tnote{****} \\ $(sec^{-1})$ } &  \shortstack{system \\ number}  \\
\hline \hline
1.0  &  50.0  & 1000 & $T_{187}= 6.6544 \times 10^{22}$ & $T_{188}= 4.6654 \times 10^{27}$ & $ T_{187+188} = 4.6655 \times 10^{27}$  & system (91) \\  
  &  & 100                 & $T_{189}= 6.6544 \times 10^{21}$ & $T_{190}= 4.6654 \times 10^{26}$ & $ T_{189+190} = 4.6655 \times 10^{26} $ & system (92) \\  
  &  & 10                   & $T_{191}= 6.6544 \times 10^{20}$ & $T_{192}= 4.6654 \times 10^{25}$ & $ T_{191+192} = 4.6655 \times 10^{25} $ & system (93) \\  
  &  & 1                     & $T_{193}= 6.6544 \times 10^{19}$ & $T_{194}= 4.6654 \times 10^{24}$ & $ T_{193+194} = 4.6655 \times 10^{24} $ & system (94) \\  \hline

50.0  &  1.0  & 1000 & $T_{195}= 8.2445 \times 10^{26}$ & $T_{196}= 3.8938 \times 10^{23}$  & $ T_{195+196} = 8.2484 \times 10^{26} $ & system (95) \\ 
 & & 100                   & $T_{197}= 8.2445 \times 10^{25}$ & $T_{198}= 3.8938\times 10^{22}$   & $ T_{197+198} = 8.2484 \times 10^{25} $ & system (96) \\ 
 & & 10                     & $T_{199}= 8.2445 \times 10^{24}$ & $T_{200}= 3.8938 \times 10^{21}$  & $ T_{199+200} = 8.2484 \times 10^{24} $ & system (97) \\ 
 & & 1                       & $T_{201}= 8.2445 \times 10^{23}$ & $T_{202}= 3.8938 \times 10^{20}$  & $ T_{201+202} = 8.2484 \times 10^{23} $ & system (98) \\  \hline \hline \hline

5.0 & 50.0 & 1000  &  $T_{203}= 2.6406 \times 10^{25}$     &    $T_{204}= 4.6636 \times 10^{27}$  & $ T_{203+204} = 4.6900 \times 10^{27} $ & system (99) \\ 
 & & 100                 &    $T_{205}= 2.6406 \times 10^{24}$   &    $T_{206}= 4.6636 \times 10^{26}$  & $ T_{205+206} = 4.6900 \times 10^{26} $ & system (100) \\ 
 & & 10                   &   $T_{207}= 2.6406 \times 10^{23}$    &    $T_{208}= 4.6636 \times 10^{25}$  & $ T_{207+208} = 4.6900 \times 10^{25} $ & system (101) \\ 
 & & 1                     &  $T_{209}= 2.6406 \times 10^{22}$     &    $T_{210}= 4.6636 \times 10^{24}$  & $ T_{209+210} = 4.6900 \times 10^{24} $ & system (102) \\  \hline

50.0 & 5.0 & 1000 &  $T_{211}= 7.1942 \times 10^{26}$     &    $T_{212}= 1.7520 \times 10^{26}$  & $ T_{211+212} = 8.9462 \times 10^{26} $ & system (103) \\ 
 & & 100                &    $T_{213}= 7.1942 \times 10^{25}$   &    $T_{214}= 1.7520 \times 10^{25}$  & $ T_{213+214} = 8.9462 \times 10^{25} $ & system (104) \\ 
 & & 10                  &   $T_{215}= 7.1942 \times 10^{24}$    &    $T_{216}= 1.7520 \times 10^{24}$  & $ T_{215+216} = 8.9462 \times 10^{24} $ & system (105) \\ 
 & & 1                    &  $T_{217}= 7.1942 \times 10^{23}$     &    $T_{218}= 1.7520 \times 10^{23}$  & $ T_{217+218} = 8.9462 \times 10^{23} $ & system (106) \\  \hline \hline \hline

1.0 & 5.0 & 1000  &  $T_{219}= 3.1768 \times 10^{23}$     &    $T_{220}= 1.7603 \times 10^{26}$  & $ T_{219+220} = 1.7662 \times 10^{26} $ & system (107) \\ 
 & & 100               &    $T_{221}= 3.1768 \times 10^{22}$   &    $T_{222}= 1.7603 \times 10^{25}$  & $ T_{221+222} = 1.7662 \times 10^{25} $ & system (108) \\ 
 & & 10                 &   $T_{223}= 3.1768 \times 10^{21}$    &    $T_{224}= 1.7603 \times 10^{24}$  & $ T_{223+224} = 1.7662 \times 10^{24} $ & system (109) \\ 
 & & 1                   &  $T_{225}= 3.1768 \times 10^{20}$     &    $T_{226}= 1.7603 \times 10^{23}$  & $ T_{225+226} = 1.7662 \times 10^{23} $ & system (110) \\  \hline

5.0 & 1.0 & 1000 &  $T_{227}= 1.4320 \times 10^{26}$     &    $T_{228}= 3.9107 \times 10^{23}$  & $ T_{227+228} = 1.4359 \times 10^{26} $ & system (111) \\ 
 & & 100              &    $T_{229}= 1.4320 \times 10^{25}$   &    $T_{230}= 3.9107 \times 10^{22}$  & $ T_{229+230} = 1.4359 \times 10^{25} $ & system (112) \\ 
 & & 10                &   $T_{231}= 1.4320 \times 10^{24}$    &    $T_{232}= 3.9107 \times 10^{21}$  & $ T_{231+232} = 1.4359 \times 10^{24} $ & system (113) \\ 
 & & 1                  &  $T_{233}= 1.4320 \times 10^{23}$     &    $T_{234}= 3.9107 \times 10^{20}$  & $ T_{233+234} = 1.4359 \times 10^{23} $ & system (114) \\  \hline

\end{tabular}
\begin{tablenotes}
\item[*] DM density that sorrounds BSSs.
\item[**] CR by $ M_{1} $ star when it is in periastron and when it is in ZAMS phase.
\item[***] CR by $ M_{2} $ star when it is in apastron and when it is in ZAMS phase.
\item[****] Total CR : $ CR_{M} = CR_{M1} + CR_{M2} $
\end{tablenotes}
\end{threeparttable}
\label{table_effect_of_DM_density_b}
\end{table*}

\section{Discussion/Conclusion} \label{conclusion_discussion}
CR of DM particles in BSSs is discussed. At first, we presented the necessary equations that are needed to calculate CR by BSSs in Section \ref{theories_and_models}. Equations \ref{CR_by_hydrogen} and \ref{CR_by_hevier} are the equations that we used in MESA stellar evolutionary code to calculate CR. Equations \ref{CR_by_hydrogen} and \ref{CR_by_hevier} are functions of stars relative velocity with respect to the DM halo i.e. $ v_{\ast} $. Then, by variation of stars velocity during the elliptical motion, the amount of CR by each star will vary too. In Section \ref{binary_system_parameters}, effect of different BSS parameters on CR were investigated. The overall results of our simulations are:

\begin{itemize}
\item CR can be negative in some configurations. It means, stars lose DM instead of capturing them. This happens in stars which their relative velocity (with respect to the DM halo) is higher than their escape velocity from the surface of the stars: $ v_{\ast} > v_{esca} $ (see Section \ref{Effect_of_semi_major_axis} for more details).

\item When stars are in apastron, they capture more DM particles in comparison to the time when they are in periastron (see Section \ref{Effect_of_stellar_masses} for more details).

\item The more the total mass of a BSS is ($ M=M{1}+M_{2} $) then, the more the CR variation is (and not the CR alone) (see Section \ref{Effect_of_stellar_masses} for more details). And using Equation \ref{Equa_L_x} we can say, the more the CR variation of a system is then the more the luminosity-variation of the system is. Then, finding DM effects in high-mass BSSs in easier than finding them in low-mass ones.

\item By increasing semi-major axis, the total CR increases too (see Section \ref{Effect_of_semi_major_axis} for more details).

\item The more the eccentricity of a systems is then, the more the CR variation is (see Section \ref{Effect_of_eccentricity} for more details). So, using Equation \ref{Equa_L_x} we can say, the more the CR variation in a system is then the more the luminosity-variation in this system is too. So, and according to our simulations, DM effects boosted when stars follow elliptical rather than circular orbits.

\item The more the density of DM around a BSS is then the more the total CR by the systems is. The incresean of the total CR by the systems has linear relation with DM density that sorrounds the BSSs.

\end{itemize}

If DM particles annihilate inside stars then, they can act as a new source of energy inside stars. As CR vary periodically during the orbital motion of BSSs components, this new source of energy causes periodic luminosity-variations in BSSs. In addition, CR variation can be translated into the neutrino flux variation, as stars (like the sun \citep{2017JCAP...07..021C, 2017PhRvD..95d3007B, 2016PhRvD..94f3512M, 2016JCAP...01..039G, 2015arXiv151103500P}) are the source of neutrino emissions. These observational considerations are of particular importance for binaries that are located in the high DM density environments (e.g. near the Galactic massive black hole or regions near the center of global clusters). \\
Besides, observational evidences can be used to constrain DM properties using BSSs, which can be the subject of future studies in this respect.

\section{Acknowledgments}
Special thanks are due to Prof. Joakim Edsjö from the University of Stockholm, Sweden, and Dr. Amin Rezaei Akbarieh from the University of Tabriz, Iran, and Prof. Gianfranco Bertone from the University of Amsterdam, Netherlands, and Marco Taoso from National Institute of Nuclear Physics (INFN) Turin, Italy  for their helpful discussions during the research. We are also grateful to the anonymous referee for useful comments and suggestions that helped to improve the manuscript. Figures of this work are generated using Gnuplot 5.2.8: an interactive plotting program (URL: \url{http://www.gnuplot.info/}) and also python's visualizations library: matplotlib v3.2.1 \citep{thomas_a_caswell_2020_3714460}.

\section{Data availability}
The codes and data underlying this article are available in GitHub repository website, at \url{https://github.com/eb-hassani/Capture-rate-of-drk-matter-particles-by-stars}

\bibliographystyle{mnras}
\bibliography{My_References} 

\bsp	
\label{lastpage}
\end{document}